\documentclass[10pt,
twocolumn,
nofootinbib,
 amsmath,amssymb,
 aps,
 pra,
floatfix,
showkeys
]{revtex4-2}

\usepackage{graphicx}
\usepackage{xcolor}
\usepackage{dcolumn}
\usepackage{bm}
\usepackage{color}
\usepackage[normalem]{ulem}
\usepackage{hyperref}
\usepackage{booktabs}

\usepackage{lipsum}

\begin{document}
 
\title{Towards Gravitational Wave Turbulence within the Hadad-Zakharov metric}

\author{Beno\^it Gay}
    \email{benoit.gay@lpp.polytechnique.fr}
\affiliation{
    Laboratoire de Physique des Plasmas, Université Paris-Saclay, CNRS, École Polytechnique, Sorbonne Université, Observatoire de Paris, F-91128 Palaiseau, France
}

\author{Eugeny Babichev}
    \email{eugeny.babichev@ijclab.in2p3.fr}
    \affiliation{Universit\'e Paris Saclay, CNRS/IN2P3, IJCLab, 91405 Orsay, France}
    
\author{S\'ebastien Galtier}
\email{sebastien.galtier@lpp.polytechnique.fr}
\affiliation{
    Laboratoire de Physique des Plasmas, Université Paris-Saclay, CNRS, École Polytechnique, Sorbonne Université, Observatoire de Paris, F-91128 Palaiseau, France}
    
\author{Karim Noui}
    \email{karim.noui@ijclab.in2p3.fr}
    \affiliation{
    Universit\'e Paris Saclay, Laboratoire de Physique des deux Infinis IJCLab,  CNRS/IN2P3, France 
    }

\date{\today}

\begin{abstract}
The theory of gravitational wave turbulence describes the long-term statistical behaviour of a set of weakly nonlinear interacting waves. 
In this paper, we aim to study aspects of gravitational turbulence within the framework of general relativity using the Hadad–Zakharov (HZ) metric. The latter is parameterised by four functions (the coefficients of a diagonal metric) that must satisfy seven non-trivial Einstein equations, six of which are independent. The issue of their mutual compatibility is therefore essential, yet it has so far been overlooked. In this work, we argue that these equations can be compatible in the weakly nonlinear regime under specific conditions. Our analytical investigation is complemented by direct numerical simulations performed with a new GPU-based code, TIGER. A comparative analysis of the evolution of the Ricci and Kretschmann scalars indicates that gravitational wave turbulence corresponds to the propagation of a genuine physical degree of freedom. These numerical findings, however, must be interpreted with caution, given the difficulty of satisfying all seven Einstein equations simultaneously with sufficient accuracy.
On the other hand, our simulations reproduce well the expected properties of the wave turbulence regime, with the emergence of a dual cascade of energy and wave action, and for the latter the observation of the Kolmogorov-Zakharov spectrum. 
In addition, our analysis reveals that the canonical variables of the problem evolve towards a nearly Gaussian statistical distribution punctuated by intermittent coherent (spatially localised and long-living) structures. 
In contrast to the canonical variables, the structure functions of the gauge-invariant metric components exhibit monofractal behaviour, which is a classical property of wave turbulence.
\end{abstract}


\maketitle

\allowdisplaybreaks

\section{Introduction}

The direct detection of gravitational waves \cite{ligoandvirgoscientificcollaborations_2016}, along with recent findings from the NANOGrav survey \cite{nanogravcollaboration_2023}, offers further evidence for the existence of a stochastic gravitational wave background. These observational advances prompt renewed interest in the long-term dynamics of gravitational waves. For gravitational waves with very low amplitudes ($h \sim 10^{-21}$), a linearised treatment remains appropriate \cite{maggiore_2007}. However, various processes in the early Universe could have generated gravitational waves with significantly larger amplitudes. Notable sources of such high-amplitude waves include first-order phase transitions \cite{krauss_1992, kosowsky_1992, kamionkowski_1994}, cosmic inflation \cite{rubakov_1982, guzzetti_2016}, self-ordering dynamics of scalar fields \cite{fenu_2009}, and the evaporation \cite{Inomata_2020} or gravitational interaction \cite{Papanikolaou_2021} of primordial black holes. In such regimes, non-linear effects are expected to play a significant role and cannot be neglected. 

One possible framework for addressing the non-linear dynamics of gravitational waves is the statistical theory of wave turbulence \cite{nazarenko_2011, galtier_2022}. This theory provides a systematic set of mathematical tools for analysing the long-time statistical behaviour of an ensemble of interacting random waves. Notably, unlike eddy turbulence, wave turbulence theory is not affected by the closure problem: the presence of a small parameter—namely, the wave amplitude—allows for a natural asymptotic closure \cite{benney_1966, deng_2022}. A central result of the theory is the derivation of the kinetic equation, which governs the temporal evolution of spectral quantities associated with conserved invariants, such as energy or wave action in the present context. The resulting dynamics are characterised by scale-by-scale spectral transfer, whereby transport from large to small scales is termed a direct cascade, while transport from small to large scales is referred to as an inverse cascade. In the wave turbulence regime, the direction of the cascade can be proved, as can the stationary spectra, which are exact solutions of the kinetic equation. 

Wave turbulence theory has demonstrated its relevance across a wide range of physical systems, including oceanography \cite{benney_1962, benney_1967c, zakharov_1967a}, plasma physics \cite{sagdeev_1969, vedenov_1967, zakharov_1967, galtier_2006}, optics \cite{dyachenko_1992, laurie_2012}, quantum mechanics \cite{dyachenko_1992, laurie_2010}, and the physics of vibrating plates \cite{during_2006}. The principal mathematical distinction among these systems lies in the order of nonlinearity, which is determined by the number of interacting waves. Whereas most systems are governed by triadic wave interactions, others involve quartic interactions, for which the dynamics evolves on significantly longer timescales, of order $\sim 1/(\omega \epsilon^4)$, compared with $\sim 1/(\omega \epsilon^2)$ in the triadic case. Here, $\epsilon \ll 1$ denotes the dimensionless wave amplitude, quantifying the smallness of the perturbation, and $\omega$ is the angular frequency \cite{benney_1967c, gay_2024}. Moreover, under appropriate symmetry conditions, quartic interactions conserve two invariants that undergo cascades in opposite directions, giving rise to the so-called dual cascade phenomenon \cite{ZLF92, nazarenko_2011, galtier_2022}. To date, the most extensively studied example of quartic wave turbulence is that of surface gravity waves, for which a substantial body of theoretical and numerical work exists \cite{onorato_2002, annenkov_2006, korotkevich_2008, korotkevich_2008a, korotkevich_2012}.

The possible existence of turbulent cascades in general relativity has been the subject of increasing attention over the past decade. In particular, non-linear perturbative approaches have been developed primarily in the context of anti-de Sitter spacetimes, where scalar-field trapping can occur. Such spacetimes have been shown to be unstable under arbitrarily small generic perturbations, leading to energy transfer to higher frequencies and, in some cases, to black hole formation \cite{bizon_2011, bizon_2015, bizon_2017}. Subsequent numerical investigations have further suggested a turbulent character of spherically symmetric gravitational collapse in asymptotically anti-de Sitter space, with the scalar curvature exhibiting a power-law spectrum associated with the  Kolmogorov-Zakharov spectrum \cite{deoliveira_2013}. However, it should be noted that in the field of turbulence, the term “Kolmogorov–Zakharov spectrum” is reserved for the exact solution of the kinetic equation of weak wave turbulence.
It should also be emphasised that conclusions regarding instability are not universal: alternative configurations have been shown to remain non-linearly stable \cite{dias_2012, dias_2012a}. More recently, studies based on simplified models of gravitational turbulence have examined families of spacetimes with stable light rings, providing evidence for the existence of a direct energy cascade that does not necessarily trigger instabilities \cite{redondo-yuste_2025}. Two-dimensional numerical simulations within the gravity–fluid correspondence support the presence of an inverse energy cascade, whose origin can be traced to the system's dimensionality, in close analogy to two-dimensional hydrodynamic turbulence \cite{carrasco_2012, green_2014}. Indication of an inverse cascade has also recently been found in a numerical simulation of the black hole environment \cite{ma_2025}. It should be noted, however, that in the field of turbulence, the concept of an “inverse cascade” requires access to a continuum of modes \cite{nazarenko_2011, galtier_2022}, a situation that remains difficult to achieve in general relativity, where only a few modes are excited. 

Interesting though they are, none of the previous studies is based on a solid theoretical framework from which precise properties of turbulence can be rigorously deduced. The first advances in this direction were made with the analytical theory of gravitational wave turbulence \cite{galtier_2017} upon a Minkowski space-time. This study shows that triadic interactions are absent and that the kinetic equation involves quartic interactions. It is shown that, due to the symmetry of the kinetic equation, the total wave action is a conserved quantity, implying that gravitational wave turbulence is characterised by a dual cascade, with an inverse cascade of wave action and a direct cascade of energy. In the stationary case, the corresponding (one-dimensional) Kolmogorov-Zakharov spectra are $k^{-2/3}$ and $k^0$, respectively. These spectra are exact solutions of the kinetic equation. The theory also predicts that the inverse cascade of wave action is explosive, in the sense that it can reach the smallest mode within a finite time (see \cite{galtier_2019} for a numerical illustration).
However, as the inverse cascade develops, the assumptions underpinning weak turbulence theory eventually break down due to the growth of nonlinearity, up to a critical scale $k_s$ at which a critical balance between linear and non-linear terms is expected to emerge. Below this scale, the wave-action spectrum may transition to a $k^{-1}$ power law, in agreement with phenomenological arguments. This transition remains compatible with the explosive character of the inverse cascade, which could ultimately lead to the formation of a condensate at low wavenumbers with potential application to the early universe \cite{galtier_2020,clough_2018}. 
An initial campaign of direct numerical simulations at moderate resolution has successfully observed the dual cascade within the validity limits of the weak turbulence regime \cite{galtier_2021}. Taken together, these results provide strong evidence for the existence of a gravitational wave turbulence regime.

The aim of the present article is twofold. 
First (Section 2), we study essential theoretical aspects of the Hadad–Zakharov metric \cite{hadad_2014} and investigate its physical relevance. In particular, we show that it admits an interpretation in three-dimensional general relativity coupled to a scalar field, thereby making manifest that the system propagates a single dynamical degree of freedom. Returning to the four-dimensional framework, we demonstrate that this degree of freedom corresponds to the “+” polarisation mode when considering linear perturbations around Minkowski spacetime.
We also identify and discuss potential issues in satisfying the full set of Einstein equations within our framework.
Second (Sections 3 and 4), we present the new numerical code and then extend the numerical analysis initiated in  \cite{galtier_2021} by investigating a range of statistically relevant quantities in both wave turbulence and general relativity. These advances are enabled by graphics processing units (GPU), whose computational power substantially reduces the time required for the direct numerical simulation of the equations.
Finally, a conclusion is given in Section 5. 

Throughout this paper, we adopt units in which the speed of light, $c$, is set to unity. In this system, the dispersion relation for gravitational waves reduces to $\omega_{\boldsymbol{k}} = k$, where $k = |\boldsymbol{k}|$ denotes the magnitude of the wave vector. We will also use the convenient notations $\alpha_{x}=\partial_x \alpha$, $\alpha_{xy}=\partial_x \partial_y \alpha$ and so on, for partial derivatives of any functions $\alpha$. 
We will assume the universe is empty, and we will not introduce a cosmological constant.
\section{Theoretical context}
This section is devoted to the investigation of several theoretical aspects of the Hadad–Zakharov metric. We begin by recalling the explicit form of the metric and deriving the associated Einstein equations. We then show that it admits an interpretation in three-dimensional general relativity coupled to a scalar field, thereby making manifest that the system propagates a single dynamical degree of freedom.

Returning to the four-dimensional framework, we demonstrate that this degree of freedom corresponds to the $+$ polarisation mode when considering linear perturbations around Minkowski spacetime. Finally, we analyse the structure and consistency of the equations in the weak turbulence regime.

\subsection{The Hadad-Zakharov ansatz}
\label{subsec:HZ}

When looking for specific solutions of Einstein's equations in vacuum, Yaron Hadad and Vladimir Zakharov considered a diagonal metric with one spatial Killing vector field in a Cartesian-like coordinate system $(t, x, y, z)$ in \cite{hadad_2014}. 
For the sake of simplicity, they chose $\partial_z$ to be the Killing vector (i.e. $\partial_z g_{\mu \nu}=0$). Under these assumptions, they proposed the following convenient parametrisation of the metric:
\begin{equation}
    g_{\mu \nu} =
    \begin{pmatrix}
        -\gamma^2 \mathrm{e}^{- 2\phi} & 0 & 0 & 0 \\
        0 & \beta^2 \mathrm{e}^{- 2\phi} & 0 & 0 \\
        0 & 0 & \alpha^2 \mathrm{e}^{- 2\phi} & 0 \\
        0 & 0 & 0 & \mathrm{e}^{2 \phi} \\
    \end{pmatrix},
\label{eq:hadad_zakharov_metric}
\end{equation}
where, $\alpha$, $\beta$, $\gamma$ and $\phi$ are four functions of $t$, $x$ and $y$.

Introducing this \textit{Ansatz} into the Einstein equations in vacuum (i.e. $R_{\mu\nu}=0$ where $R_{\mu\nu}$ is the Ricci tensor), one finds a set of seven non-trivial equations. The three off-diagonal components of the Einstein equations $R_{tx}=0$, $R_{ty}=0$ and $R_{xy}=0$ lead, correspondingly
\begin{subequations}
\begin{align}
     \alpha_{tx} &=
    - 2 \alpha \,  \phi_t  \phi_x
    + \frac{ \beta_t  \alpha_x}{\beta}
    + \frac{ \alpha_t  \gamma_x}{\gamma} \, ,
    \label{eq:constraint_alpha} 
    \\
     \beta_{ty} &=
    - 2 \beta \,  \phi_t  \phi_y
    + \frac{ \alpha_t  \beta_y}{\alpha}
    + \frac{ \beta_t  \gamma_y}{\gamma} \, ,
    \label{eq:constraint_beta}
    \\
     \gamma_{xy} &=
    - 2 \gamma \,  \phi_x  \phi_y
    + \frac{\alpha_x  \gamma_y}{\alpha}
    + \frac{ \gamma_x  \beta_y}{\beta}
    ,
    \label{eq:constraint_gamma}
    \end{align}
while the three diagonal equations, $R_{tt}=0$, $R_{xx}=0$ and $R_{yy}=0$, give, correspondingly
\begin{align}
   \label{eq:G_zz}
    \alpha \, 
    \alpha_{xx}  &+ \beta \,  \beta_{yy} 
    = \\\notag
    & - \frac{\alpha^2 \beta^2}{\gamma^2}  \phi_t ^2
    - \alpha^2   \phi_x 
    - \beta^2   \phi_y  \\\notag
    & + \frac{\alpha \beta}{\gamma}  \alpha_t  \beta_t
    + \frac{\alpha}{\beta}  \alpha_x \beta_x
    + \frac{\beta}{\alpha}  \alpha_y  \beta_y,
    \\
    \label{eq:G_yy}
    \beta \,  \beta_{tt} & - \gamma_{xx} \,  \gamma 
    = \\\notag
    & - \beta^2   \phi_t 
    + \gamma^2   \phi_x 
    - \frac{\beta^2 \gamma^2}{\alpha^2}   \phi_y  \\\notag
    & + \frac{\beta}{\gamma}  \beta_t  \gamma_t
    - \frac{\gamma}{\beta} \beta_x  \gamma_x
    + \frac{\beta \gamma}{\alpha^2}  \beta_y  \gamma_y,
    \\
    \label{eq:G_xx}
    \alpha \,  \alpha_{tt} & - \gamma\, \gamma_{yy}
    = \\\notag
    & - \alpha^2   \phi_t 
    - \frac{\alpha^2 \gamma^2}{\beta^2}   \phi_x 
    + \gamma^2  \phi_y  \\\notag
    & + \frac{\alpha}{\gamma}  \alpha_t  \gamma_t
    + \frac{\alpha \gamma}{\beta^2}  \alpha_x  \gamma_x
    - \frac{\gamma}{\alpha}  \alpha_y  \gamma_y.
\end{align}
Finally, instead of the $zz$ component of the Einstein equations, one can conveniently take the combination $R-2 R^z_z=0$, where $R$ is the Ricci scalar, yielding the dynamical equation on $\phi$,
\begin{equation}
    \partial_t \left( \frac{\alpha \beta}{\gamma}  \phi_t \right)
    - \partial_x \left( \frac{\alpha \gamma}{\beta}  \phi_x \right)
    - \partial_y \left( \frac{\beta \gamma}{\alpha}  \phi_y \right)
    = 0.
    \label{eq:dynamic} 
\end{equation}
\label{eq:systeme_initial}
\end{subequations}

The resulting system consists of seven equations for four unknown fields, making it overdetermined and raising the possibility of inconsistency. Acknowledging this concern, Hadad and Zakharov stated in their original paper that the system is "compatible." The compatibility of this system of equations is linked to the Bianchi identities. In fact, as we will demonstrate below, there are three non-trivial Bianchi identities associated with the ansatz~(\ref{eq:hadad_zakharov_metric}). This relationship naively suggests that we have the correct count of free functions: seven independent equations minus three Bianchi identities should correspond to four independent functions. However, an important nuance to consider is that the Bianchi identities do not automatically guarantee that the diagonal equations~(\ref{eq:G_xx}–\ref{eq:G_zz}) will be satisfied once the off-diagonal ones (\ref{eq:constraint_alpha}–\ref{eq:constraint_gamma}) and the dynamical equation for $\phi$ (\ref{eq:dynamic}) are satisfied.

Indeed, the 3 non-trivial Bianchi identities, once (\ref{eq:constraint_alpha}-\ref{eq:constraint_gamma}) are taken into account, can be reformulated as follows, 
\begin{subequations}
\label{Bianchi}
\begin{align}
&&\gamma \, \partial_t ( A R_{tt}) + \beta_t (B R_{xx}) + \alpha_t (C R_{yy}) \, = \,  0 \, , \label{BI0} \\
&&\beta \, \partial_x ( B R_{xx}) + \gamma_x (A R_{tt}) - \alpha_x (C R_{yy}) \, = \,  0 \, , \label{BI1} \\
&&\alpha \, \partial_y ( C R_{yy}) + \gamma_y (A R_{tt}) - \beta_y (A R_{xx}) \, = \,  0 \, , \label{BI2} 
\end{align}
\end{subequations}
where we have introduced shorthand notations for simplicity
\begin{eqnarray}
A=\frac{\alpha \beta}{\gamma^2} \, , \quad
B=\frac{\alpha \gamma}{\beta^2} \, , \quad
C=\frac{\beta \gamma}{\alpha^2} \, .
\end{eqnarray}
As one can see, the above equations allow the trivial solutions $R_{tt}=R_{xx}=R_{yy} =0$, which would imply that all the Einstein equations were satisfied once Eqs.~(\ref{eq:constraint_alpha},\ref{eq:constraint_beta},\ref{eq:constraint_gamma}) are solved. In this sense, the system of equations is indeed compatible. 
To assure this, one must impose that 
\begin{equation}
\label{good_ic}
R_{tt} = R_{xx}=R_{yy} =0 \quad \text{at $t=0$} \,.
\end{equation}
On the other hand, if the initial conditions are such that Eq.~(\ref{good_ic}) is not satisfied, one does not expect that the equations $R_{tt} = R_{xx}=R_{yy} =0$ are satisfied during the evolution, thus making the system of equation to be a theory that does not describe General Relativity. 

Before studying the resolution of these field equations, we first examine some geometrical properties of the Hadad-Zakharov space-time in order to gain a clearer physical interpretation. 

\subsection{Reformulation as a three-dimensional theory of gravity minimally coupled to a scalar}

In this section, we interpret the Hadad-Zakharov metric in the framework of 2+1-dimensional gravity via dimensional reduction. 
For that purpose, we start considering a four-dimensional metric  $g_{\mu \nu}$  whose components, expressed in Cartesian coordinates $(t,x,y,z)$ are independent of the coordinate  $z$ and satisfy:
\begin{equation}
    g_{tz} = g_{xz} = g_{yz} = 0 \quad  \text{and} \quad g_{zz}= e^{2 \phi}.
\end{equation}
All remaining components are left unrestricted at this stage. The four-dimensional line element may then be written as:
\begin{equation}
    \mathrm{d}s^2 = \hat{g}_{\mu\nu} \mathrm{d}\hat{x}^\mu \mathrm{d}\hat{x}^\nu + \mathrm{e}^{2\phi} \mathrm{d}z^2,
\label{eq:line_element}
\end{equation}
where $\hat{g}$ denotes the induced three-dimensional metric on the manifold parametrised by the coordinates $\hat{x}^{\mu}=(t,x,y)$.

A direct computation shows that, for metrics of the form \eqref{eq:line_element}, the four-dimensional Ricci scalar $R$ reduces to the three-dimensional Ricci scalar $\hat{R}$. Consequently, the four-dimensional Einstein–Hilbert action reduces to the following three-dimensional action:
\begin{equation}
    \hat{S}[\hat{g},\phi] = \int \mathrm{d}^3x \sqrt{\det \hat{g}} \;  \mathrm{e}^{\phi} \; \hat{R},
    \label{eq:3Daction}
\end{equation}
where the integral over the $z$ variable has been removed. It follows that, for the class of metrics satisfying $\partial_z g_{\mu \nu} = 0$ with $\mu$, $\nu \in \{t,x,y,z\}$ and the decomposition \eqref{eq:line_element}, the four-dimensional Einstein–Hilbert action reduces to a three-dimensional scalar–tensor theory with a non-minimal coupling with the scalar field $\phi$. Such a theory admits a single propagating physical degree of freedom, since pure three-dimensional general relativity is well known to be a topological field theory with no local degrees of freedom. This degree of freedom is therefore associated with the presence of the scalar field $\phi$. 

The three-dimensional action can be further simplified by performing a conformal transformation of the metric, namely:
\begin{equation}
    \tilde{g}_{\mu\nu} = e^{2 \phi} \hat{g}_{\mu\nu}.
\end{equation}
For such a metric, one readily finds that the action \eqref{eq:3Daction} is mapped onto a standard scalar--tensor theory with a minimally coupled scalar field:
\begin{equation}
    \hat{S}[\tilde{g},\phi] = \int d^3x \, \sqrt{\det{\tilde{g}}} \, \left(  \tilde{R} \, - 2 \tilde{g}^{\mu\nu} \, \phi_\mu \,   \phi_\nu \right),
    \label{eq:3Daction2}
\end{equation}
where $\tilde{R}$ denotes the Ricci scalar associated with the metric $\tilde{g}_{\mu\nu}$. This conformal reformulation makes it manifest that the theory propagates a single physical degree of freedom encapsulated in $\phi$.

We now turn to the equations of motion. The six equations governing the metric components are obtained from
\begin{equation}
    \mathcal{E}_{\mu\nu} \;=\; \tilde{G}_{\mu\nu} - T_{\mu\nu} \;=\; 0 ,
\end{equation}
where $\tilde{G}_{\mu\nu}$ denotes the Einstein tensor associated with the metric $\tilde{g}_{\mu\nu}$, and $T_{\mu\nu}$ the stress-energy tensor of the minimally coupled scalar field. It is given by:
\begin{equation}
   T_{\mu\nu} \equiv \tilde{X} \tilde{g}_{\mu\nu}
   + 2 \,  \phi_\mu \,  \phi_\nu ,
   \label{eq:eomRicci}
\end{equation}
where the first term $\tilde{X}=-\tilde{g}^{\alpha\beta}  \phi_\alpha   \phi_\beta$ corresponds to the kinetic energy density of the scalar field. Hence, the equation of motion for the scalar field reads as follows:
\begin{eqnarray}
    \mathcal{E}_\phi & \equiv & \tilde{\Box} \phi
    \; \equiv \; \tilde{g}^{\alpha \beta} \tilde{\nabla}_\alpha \tilde{\nabla}_\beta \phi \\
    & = & \frac{1}{\sqrt{\det \tilde g}} \,
    \partial_\alpha \left( \sqrt{\det \tilde g} \, \tilde{g}^{\alpha \beta}  \phi_\beta \right)
    \; = \; 0 ,
    \label{eq:eomphi}
\end{eqnarray}
where $\tilde{\nabla}_\alpha$ denotes the covariant derivative compatible with the metric $\tilde{g}_{\mu\nu}$.

However, diffeomorphism invariance implies that the scalar-field equation of motion is not independent, but can instead be derived from the metric equations. Indeed, invoking the Bianchi identity: $\tilde{\nabla}^\mu \tilde{G}_{\mu\nu} = 0$, one readily finds that taking the covariant divergence of \eqref{eq:eomRicci} yields the scalar-field equation \eqref{eq:eomphi}, namely:
\begin{equation}
    \tilde{\nabla}^\nu \mathcal{E}_{\mu\nu} \;=\; 2 \, \mathcal{E}_\phi \, \phi_\mu \, .
    \label{Bianchi3D}
\end{equation}
As a consequence, only six of the equations of motion are independent.

From the Hamiltonian point of view, it is straightforward to see that the theory admits only one physical degree of freedom. Indeed, one introduces a (local) foliation ${\cal M}=\Sigma \times \mathbb R$  of the 3-dimensional space $\cal M$ into 2-dimensional surfaces $\Sigma$, and one considers the usual ADM parametrization of the metric
\begin{eqnarray}
    d\tilde{s}^2 = -N^2 dt^2 + \tilde{\gamma}_{ij}(dx^i+ N^i dt)(dx^j+ N^j dt) \, ,
    \end{eqnarray}
where $N$ is the lapse function, $N^i$ the two components of the shift vector and $\gamma_{ij}$ the three components of the 2-dimensional  induced metric on the surface $\Sigma$. It is well-known that $N$ and $N^i$ are Lagrange multipliers respectively associated to the Hamiltonian constraint and the two vectorial constraints. Furthermore, the Hamiltonian constraint corresponds to the equation ${\cal E}^{00}=0$ while the remaining two constraints are given by ${\cal E}^{0i}=0$. These three constraints are first class: they generate infinitesimal diffeomorphisms on phase space; their Poisson brackets close among themselves; the resulting algebra reproduces the algebra of three-dimensional diffeomorphisms. In other words, they implement the diffeomorphism symmetry of the theory at the level of phase space. As a consequence, once these constraints are satisfied at the initial time, they remain preserved throughout the evolution. 
Finally, as the phase space is parametrized by the 3 components of the spatial metric $\tilde{\gamma}_{ij}$, the scalar field $\phi$ and their associated momenta, the theory admits only one degree of freedom ($4$ variables - $3$ first class constraints), as we said above. Notice the remaining three Einstein equations ${\cal E}^{ij}=0$ are recovered in the Hamiltonian context from the time evolution of the three components $\tilde{\gamma}_{ij}$ and their momenta.

Following \cite{hadad_2014}, we now restrict attention to diagonal metrics of the form:
\begin{equation}
    \tilde{g}_{\mu\nu} = \mathrm{diag}(-\gamma^2, \beta^2, \alpha^2),
\end{equation}
where $\alpha$, $\beta$ and $\gamma$ are positive functions. It is straightforward to verify that the three off-diagonal equations, ${\cal E}_{tx}$, ${\cal E}_{ty}$  and ${\cal E}_{xy}$ are respectively equivalent to  \eqref{eq:constraint_alpha}, \eqref{eq:constraint_beta} and \eqref{eq:constraint_gamma}, hence the first two, \eqref{eq:constraint_alpha} and \eqref{eq:constraint_beta}, are the vectorial constraints. The third equation \eqref{eq:constraint_gamma} should be the evolution equation of the component $\gamma_{xy}$, however, as $\gamma_{xy}=0$ in the HZ ansatz, the equation looks like a constraint. 

As for the three diagonal equations ${\cal E}_{\mu\mu}=0$, they coincide with the remaining equations obtained from the diagonal components of the four-dimensional Ricci tensor given in  \cite{hadad_2014}. Notice that \eqref{eq:G_zz} is the Hamiltonian constraint.

Finally, the scalar field equation \eqref{eq:eomphi} reduces to:
\begin{equation}
   -\partial_t\left(\frac{\alpha \beta}{\gamma} \,  \phi_t \right)
   + \partial_x \left(\frac{\alpha \gamma}{\beta} \, \phi_x \right)
   + \partial_y \left(\frac{\beta \gamma}{\alpha} \,  \phi_y \right)
   = 0,
\end{equation}
which corresponds to equation (21) in \cite{hadad_2014}.

Concerning the 3-dimensional Bianchi identities \eqref{Bianchi3D}, one can immediately see that they reproduce \eqref{BI0}, \eqref{BI1} and \eqref{BI2} as expected.

\subsection{Scalar-vector-Tensor decomposition of the Hadad-Zakharov metric}

In this Section, we return to the four-dimensional formulation and study small perturbations around the Minkowski space-time
$\eta_{\mu\nu} = \mathrm{diag}(-1,1,1,1)$ of the Hadad–Zakharov metric~\eqref{eq:hadad_zakharov_metric} with the aim of gaining clearer physical insight into the propagating degrees of freedom encoded in the metric in the weak field regime. 

Assuming that deviations from the Minkowski metric are small, the spacetime metric may be written as
$g_{\mu\nu} = \eta_{\mu\nu} + h_{\mu\nu}$,
with $h_{\mu\nu} \ll 1$.
For the Hadad–Zakharov ansatz, the metric perturbation takes the diagonal form
\begin{equation}
    h_{\mu \nu} = \mathrm{diag}\!\left(h_{tt}, h_{xx}, h_{yy}, h_{zz}\right),
    \label{eq:HZpert}
\end{equation}
with components given by:
\begin{subequations}
\begin{align}
    h_{tt} &= - \gamma^2 \mathrm{e}^{-2\phi} + 1, \label{HZtt} \\
    h_{xx} &= \beta^2 \mathrm{e}^{-2\phi} - 1, \label{HZxx} \\
    h_{yy} &= \alpha^2 \mathrm{e}^{-2\phi} - 1, \label{HZyy} \\
    h_{zz} &= \mathrm{e}^{2\phi} - 1. \label{HZzz}
\end{align}
\end{subequations}
These expressions should be compared with the standard decomposition of metric perturbations into scalar, vector, and tensor modes (the SVT decomposition). This decomposition may be written as follows (see \cite{Mukhanov_2005} for example):
\begin{subequations}
\begin{align}
    h_{00} &= -2 \Phi, \label{h00}\\
    h_{0i} &= \partial_i B + S_i, \label{h0i}\\
    h_{ij} &= -2 \Psi \, \delta_{ij}
    + 2 \partial_{ij} E
    + \partial_j F_{i} + \partial_i F_{j}
    + h_{ij}^{TT}, \label{hij}
\end{align}
\end{subequations}
where $\Phi$, $B$, $E$ and $\Psi$ are scalars under three-dimensional spatial rotations, whereas $S_i$ and $F_i$ transform as vectors.
Note that Latin indices $i, j$ correspond to three-dimensional coordinates $(x, y, z)$. Furthermore, the following constraints are imposed: 
\begin{align}
   \partial_i S_{i}=0 \, , \quad
   \partial_i F_{i}=0 \, , \quad
   h_{ii}^{TT}=0 \, , \quad
   h^{TT}_{i j, j}=0 \, . 
   \end{align}
The dynamical gravitational degrees of freedom are encoded in the traceless and transverse 3-tensor $h_{i j}^{T T}$, which is gauge invariant.  

Since the Hadad–Zakharov ansatz is independent of the coordinate $z$, we consistently assume that all SVT variables depend only on $t$, $x$ and $y$. By comparing the equations (\ref{HZtt}) and (\ref{h00}), one immediately obtains:
\begin{equation}
    \Phi = - \frac12 h_{tt},
    \label{Phi}
\end{equation}
while comparison of the mixed components $h_{0i}$ yields:
\begin{equation}
    B=0, \quad S_i =0.
    \label{BS}
\end{equation}
Next, using the tracelessness and divergence-free of $h^{TT}_{ij}$, one may write:
\begin{align}
    h_{i i} &= h_{x x}+h_{y y}+h_{z z}
    = -6 \Psi + 2 \Delta E, \\
    \partial_{ij} h_{ij} &= \partial_{xx} h_{xx} + \partial_{yy} h_{yy}
    = -2 \Delta \Psi + 2 \Delta^2 E,
\end{align}
where $\Delta = \partial_x^2 + \partial_y^2$ denotes the two-dimensional Laplacian. Combining these relations allows the two scalar gravitational potentials to be expressed as:
\begin{align}
    \Psi &= -\frac{1}{4} h_{i i} + \frac{1}{4} \Delta^{-1} \partial_{ij} h_{ij}, \\
    E &= -\frac{1}{4} \Delta^{-1} h_{i i} + \frac{3}{4} \Delta^{-2} \partial_{ij} h_{ij},
    \label{PsiE}
\end{align}
where $\Delta^{-1}$ denotes the inverse Laplacian operator.

Using equation \eqref{PsiE} together with the divergence of the spatial components of the metric,
\begin{equation}
    \partial_{j} h_{ij}
    = -2 \partial_i \Psi + 2 \Delta \partial_i E + \Delta F_i,
\end{equation}
one obtains:
\begin{equation}
    F_i
    = \Delta^{-1} \partial_j h_{ij}
    - \Delta^{-2} \partial_{ijk} h_{jk} .
\end{equation}
This relation may be written explicitly in component form as:
\begin{subequations}
\begin{align}
    F_x &= \Delta^{-1} \, \partial_x
    \left( h_{xx} - \Delta^{-1} \partial_{jk} h_{jk} \right), \\
    F_y &= \Delta^{-1} \, \partial_y
    \left( h_{yy} - \Delta^{-1} \partial_{jk} h_{jk} \right), \\
    F_z &= 0.
    \label{F}
\end{align}
\end{subequations}

Having expressed all scalars and vector potentials in terms of the components of the Hadad–Zakharov metric, we may now determine the transverse–traceless part of the metric perturbation. By comparing Eq.~\eqref{hij} with equations (\ref{HZxx})–(\ref{HZzz}), and making use of equations \eqref{Phi}, \eqref{BS}, \eqref{PsiE}, and \eqref{F}, one obtains the non-vanishing components of the transverse–traceless metric perturbation:
\begin{subequations}
\begin{align}
    h_{xx}^{TT} &=
    h_{xx}
    - \frac{1}{2} h_{ii}
    + \frac{1}{2} \Delta^{-1} \partial_{ij} h_{ij} \\\notag
    & \hspace{50pt}
    + {\Delta}^{-1}{\partial_x^2}
    \left(
        \frac{1}{2} h_{ii}
        - 2 h_{xx}
        + \frac{1}{2} \Delta^{-1} \partial_{ij} h_{ij}
    \right), \\
    h_{yy}^{TT} &=
    h_{yy}
    - \frac{1}{2} h_{ii}
    + \frac{1}{2} \Delta^{-1} \partial_{ij} h_{ij} \\\notag
    & \hspace{50pt}
    + {\Delta}^{-1}{\partial_y^2}
    \left(
        \frac{1}{2} h_{ii}
        - 2 h_{yy}
        + \frac{1}{2} \Delta^{-1} \partial_{ij} h_{ij}
    \right), \\
    h_{xy}^{TT} &=
    {\Delta}^{-1}{\partial_x \partial_y}
    \left(
        \frac{1}{2} h_{ii}
        - h_{xx}
        - h_{yy}
        + \frac{1}{2} \Delta^{-1} \partial_{ij} h_{ij}
    \right), \\
    h_{zz}^{TT} &=
    h_{zz}
    - \frac{1}{2} h_{ii}
    + \frac{1}{2} \Delta^{-1} \partial_{ij} h_{ij}.
\end{align}
\label{hTT}
\end{subequations}

The above expressions for the SVT potentials simplify considerably when an expansion of the Hadad-Zakharov functions $\alpha,\beta,\gamma$ and $\phi$ in a small parameter $\epsilon$ is performed. It is clear that $\phi = {\cal O}(\epsilon)$ while, as we are going to show explicitly below,
\begin{eqnarray}
\alpha = 1 +{\cal O}(\epsilon^2) \, , \quad
\beta = 1 +{\cal O}(\epsilon^2) \, , \quad
\gamma = 1 +{\cal O}(\epsilon^2) \, .
\end{eqnarray}
Hence, working at first order in $\epsilon$, one finds that the metric perturbation depends solely on the scalar field $\phi$:
\begin{equation}
    h_{\mu \nu} =
    2 \phi \, \mathrm{diag}( 1, - 1, -  1,  1 ) \, .
    \label{HZsmall}
\end{equation}
Taking equation (\ref{HZsmall}) into account, it is straightforward to derive from equations (\ref{Phi}), (\ref{BS}), (\ref{PsiE}), and (\ref{F}) the scalar and vector components at first order in $\epsilon$:
\begin{equation}
\begin{split}
\Phi &= -\phi, \quad \Psi = 0, \quad B = 0, \quad E = -\Delta^{-1} \phi , \\
S_i & = 0, \quad F_i = 0 .
\end{split}
\label{SVe}
\end{equation}

The transverse–traceless part of the metric perturbation can be obtained by expanding (\ref{hTT}) to first order in $\epsilon$ and one finds
\begin{equation}
    h_{i j}^{T T}(t,x,y) =
    \begin{pmatrix}
    - 2 \partial_{xx} E - 2 \phi & - 2 \partial_{xy} E & 0 \\
    - 2 \partial_{xy} E & -2 \partial_{yy} E - 2 \phi & 0 \\
    0 & 0 & 2 \phi
    \end{pmatrix}.
\label{hTTe}
\end{equation}
At this stage, it is appropriate to make several remarks.

First, we note that the Hadad–Zakharov metric in the weak-field regime corresponds neither to the transverse (conformal Newtonian) gauge—defined by the conditions $B = E = 0$—nor to the synchronous gauge, for which $h_{0\mu} = 0$. 

As a second remark, we observe that although the Hadad–Zakharov metric is diagonal, as well as its weak-field approximation \eqref{HZsmall}, the transverse–traceless part of the perturbation $h_{ij}^{TT}$ contains non-diagonal components, in particular $h_{xy}^{TT} = h_{yx}^{TT}$. We emphasise that this is a gauge-invariant statement, since $h_{ij}^{TT}$ itself is gauge invariant. The origin of this non-diagonal component is discussed below. 

Finally, the metric \eqref{HZsmall} contains a single dynamical degree of freedom. This property is directly inherited from the full metric \eqref{eq:hadad_zakharov_metric}. It is nevertheless instructive to recover this feature by analysing the dynamical degrees of freedom of the graviton, encoded in the transverse–traceless perturbation $h_{ij}^{TT}$, within the framework of the linearised equations.

At order $\epsilon$, the Einstein equations reduce to trivial identities with the exception of the $zz$ component, which yields:
\begin{equation}
     \phi_{tt} -  \phi_{xx} -  \phi_{yy} = 0 .
\end{equation}
This equation may equivalently be obtained by expanding equation~(\ref{eq:dynamic}) to first order in the small parameter $\epsilon$. It admits plane-wave solutions of the form
\begin{equation}
    \phi = C_0 \, \mathrm{e}^{i \omega t - i k_x x - i k_y y},
    \qquad
    \omega^2 = k_x^2 + k_y^2 ,
\end{equation}
where $C_0$ is a constant amplitude.

For such a solution, (\ref{SVe}) implies $E = \omega^{-2} \phi$, hence, the transverse–traceless part of the metric perturbation can therefore be written as:
\begin{equation}
    h_{ij}^{TT}(\omega)
    =
    \frac{2 \phi}{\omega^2}
    \begin{pmatrix}
        -k_y^2 & k_x k_y & 0 \\
        k_x k_y & -k_x^2 & 0 \\
        0 & 0 & \omega^2
    \end{pmatrix}.
    \label{homega}
\end{equation}
In the special case of a wave propagating along the $x$-direction (i.e. $k_y=0$), this expression simplifies and reduces to
\begin{equation}
\begin{aligned}
    {}^{(x)} h_{ij}^{TT}(\omega)
    &= 2 \phi
    \begin{pmatrix}
        0 & 0 & 0 \\
        0 & - 1& 0 \\
        0 & 0 & 1
    \end{pmatrix}, \\
\end{aligned}
\end{equation}
Similarly, for a wave propagating along the $y$-direction, one finds
\begin{equation}
\begin{aligned}
    {}^{(y)} h_{ij}^{TT}(\omega)
    &=2 \phi 
    \begin{pmatrix}
        -1 & 0 & 0 \\
        0 & 0 & 0 \\
        0 & 0 & 1
    \end{pmatrix}, \\
\end{aligned}
\end{equation}

It is worth emphasising that, in each case, the wave exhibits only the usual ``$+$'' polarisation, while the ``$\times$'' polarisation is absent. This observation is consistent with the fact that the theory possesses a single dynamical degree of freedom, which is precisely associated with the ``$+$'' polarisation mode.

Finally, we note that a plane-wave solution $h_{ij}^{TT}(\omega)$ corresponding to an arbitrary wave vector $\{k_x, k_y\}$ can be obtained as a simple rotation of a wave propagating along the $x$-direction in the $(x,y)$ plane.
In this way, the appearance of off-diagonal components in $h_{ij}^{TT}$ is naturally understood as a consequence of rotations in the $(x,y)$ plane.

\subsection{Weak non-linearity}

In what follows, we will be interested in weak turbulence regime, which assumes a weakly non-linear regime. In this formalism, it is convenient to quantify the amplitude of the gravitational waves using a small parameter $\epsilon$. We then consistently expand the four functions entering the HZ ansatz as follows,
\begin{equation}
\label{expansion}
\phi \rightarrow \epsilon \phi,\;\;  
\alpha \rightarrow 1+\epsilon^2 \tilde\alpha,\;\; 
\beta \rightarrow 1+\epsilon^2 \tilde\beta,\;\;  
\gamma \rightarrow 1+\epsilon^2 \tilde\gamma.
\end{equation}
With~(\ref{expansion}) taken into account, we expand the Einstein equations for HZ ansatz~(\ref{eq:systeme_initial}) up to $\epsilon^3$ order. The resulting equations take a much simpler form. 

The 3 non-trivial off-diagonal equations $tx$, $ty$, $xy$ read, correspondingly,
\begin{subequations}
\begin{align}
R_{tx}^{(3)} & =\tilde\alpha_{t x}+2 \phi_t \phi_x=0 , \label{eq:constraint_alpha_3}\\
R_{ty}^{(3)} &=\tilde\beta_{t y}+2 \phi_t \phi_y=0 , \label{eq:constraint_beta_3}\\
R_{xy}^{(3)} &=\tilde\gamma_{x y}+2 \phi_x \phi_y=0 ,
\label{eq:constraint_gamma_3}
\end{align}
where the subscripts  $(3)$ signifies that the expressions are taken at third order in $\epsilon$.

The dynamical equation~(\ref{eq:dynamic}) becomes,
\begin{align}
\label{eq:dynamic_3}
    \phi_{tt} &- \phi_{xx} - \phi_{yy} 
    =
    - \epsilon^2 \partial_t \left[ (\tilde\alpha+\tilde\beta-\tilde\gamma) \phi_t \right] \\\notag
    & 
    + \epsilon^2 \partial_x\left[(\tilde\alpha-\tilde\beta+\tilde\gamma) \phi_x \right] 
    + \epsilon^2 \partial_y\left[(\tilde\alpha+\tilde\beta+\tilde\gamma) \phi_y \right],
\end{align}
while the diagonal Einstein equations (\ref{eq:G_zz})-(\ref{eq:G_xx}) are rewritten as follows,
\begin{align}
R_{tt}^{(3)} &= \tilde\alpha_{x x}+\tilde\beta_{y y}-\left(-\phi_t^2-\phi_x^2-\phi_y^2\right)   =0, \label{eq:G_zz_3}\\
R_{xx}^{(3)} & = \tilde\alpha_{t t}-\tilde\gamma_{y y}-\left(-\phi_t^2-\phi_x^2+\phi_y^2\right)  =0, \label{eq:G_yy_3}\\
R_{yy}^{(3)} & = \tilde\beta_{t t}-\tilde\gamma_{x x}-\left(\phi_t^2+\phi_x^2-\phi_y^2\right)  =0. \label{eq:G_xx_3}
\end{align}
\label{eq:main_system}
\end{subequations}
Notice that the form of Eqs.~(\ref{eq:constraint_alpha_3})–\eqref{eq:G_xx_3} is in agreement with our expansion in small parameter~(\ref{expansion}).
Moreover, terms of order $\epsilon^3$ are absent from these equations, with the exception of the scalar-field equation \eqref{eq:dynamic_3}, where such contributions arise implicitly through non-linear couplings between $\phi$ and $\alpha$, $\beta$ or $\gamma$.

At the same order $\epsilon^3$, the Bianchi identities are especially simple, 
\begin{equation}
\label{Bianchi3}
\partial_t R_{tt}^{(3)} = 0, \quad \partial_x R_{xx}^{(3)} = 0, \quad \partial_y R_{yy}^{(3)} = 0,
\end{equation}
leading to 
\begin{equation}
\label{Bianchi_sol}
R_{tt}^{(3)} = c_0(x,y), \;\; 
R_{xx}^{(3)} = c_1(t,y), \;\; 
R_{yy}^{(3)} = c_2(t,x) \, 
\end{equation}
where $c_0$, $c_1$ and $c_2$ are in general arbitrary functions.
At this order of expansion, the question of compatibility of equations is particularly clear. 
Indeed, due to the presence of the free functions $c_0$, $c_1$ and $c_2$ in~(\ref{Bianchi_sol}), the diagonal Einstein equations are generically not satisfied even if the other 4 equations are satisfied. In order for a solution of the HZ system of equations (\ref{eq:constraint_alpha_3})-(\ref{eq:dynamic_3}) to be a solution of the Einstein equations (at this order), one additionally needs to assure that $c_0=c_1=c_2=0$.
As we will show below, this is indeed the case in the wave turbulence regime.

\subsection{A wave-turbulence description}

This section summarises the main results of \cite{galtier_2017, gay_2024}. As is standard in wave turbulence theory, the analysis is carried out in Fourier space and formulated in terms of canonical variables $a^s_{\boldsymbol{k}}(t)$, defined by:
\begin{equation}
    a^s_{\boldsymbol{k}} (t)
    =
    \left(
        \sqrt{\frac{k}{2}} \, \hat \phi_{\boldsymbol{k}}(t)
        + \frac{i s}{\sqrt{2 k}} \, \partial_t \hat \phi_{\boldsymbol{k}}( t)
    \right)
    \mathrm{e}^{i s \omega_{\boldsymbol{k}} t},
    \label{eq:canonical_variables_definition}
\end{equation}
where $s = \pm 1$ denotes the directional polarity (for a given wave vector $\boldsymbol{k}$, there are two possible propagation directions), and $\hat \phi_{\boldsymbol{k}}(t)$ is the spatial Fourier transform of $\phi(\boldsymbol{x}, t)$\footnote{Note that throughout this work, we adopt the following convention for the Fourier transform:
\begin{equation}
    \hat \phi_{\boldsymbol{k}}(t)
    = \frac{1}{(2 \pi)} \int_{\mathbb{R}^2} 
    \phi(\boldsymbol{x}, t)
    e^{-i \boldsymbol{k} \cdot \boldsymbol{x}}
    ~\mathrm{d}\boldsymbol{x}.
\end{equation}
}.
These variables are obtained by diagonalising the linearised problem in Fourier space (see, e.g., \cite{nazarenko_2011, galtier_2022}).

The canonical variables are then substituted into the system of equations \eqref{eq:main_system} in order to derive their evolution equations. At this stage, however, two difficulties arise. First, the system consists of seven equations for four unknown fields and is therefore overdetermined, requiring a choice of which subset of equations to solve. Second, substituting the canonical variables into either the diagonal or the off-diagonal equations involves time integrations, which must be handled with care.

In general, these two difficulties are non-trivial. However, within the framework of wave turbulence theory, both can be addressed by invoking a separation of timescales between the amplitude and the phase of the canonical variables. Specifically, the temporal variation of the amplitudes is assumed to be much slower than that of the oscillatory phases, which means that the linear time ($\tau_{GW} \sim 1/\omega$) is supposed to be much smaller than the non-linear time \cite{galtier_2017}. In mathematical terms, this assumption reads
\begin{equation}
    \left| \partial_t a^s_{\boldsymbol{k}} \right|
    \ll
    k \left| a^s_{\boldsymbol{k}} \right| .
\end{equation}
Under this condition, time integration can be carried out by effectively dividing by a factor of $\omega$ (or $k$). Moreover, this separation of timescales ensures the equivalence between the diagonal equations \eqref{eq:G_zz_3}–\eqref{eq:G_xx_3} and the off-diagonal constraint equations \eqref{eq:constraint_alpha_3}–\eqref{eq:constraint_gamma_3} as we demonstrate below. Note that the latter inequality means that the weak turbulence regime does not describe the $k=0$ mode (often called the slow mode), which belongs to the strong turbulence regime. Indeed, in this case, the linear time becomes infinite and cannot be slower than the non-linear time. This is a very classic situation: weak turbulence theory is always valid in a limited domain of Fourier space \cite{nazarenko_2011, galtier_2022}. 

To demonstrate the equivalence, under the assumptions of wave turbulence, between (i) the set of non-diagonal equations~(\ref{eq:constraint_alpha_3})–(\ref{eq:constraint_gamma_3}) together with the dynamical equation~(\ref{eq:dynamic_3}), and (ii) the diagonal Einstein equations~(\ref{eq:G_zz_3})–(\ref{eq:G_xx_3}) together with~(\ref{eq:dynamic_3}), we proceed as follows.
Using the diagonal equations~(\ref{eq:G_zz_3})–(\ref{eq:G_xx_3}) in conjunction with~(\ref{eq:dynamic_3}), one obtains, at $\epsilon^3$ order:
\begin{subequations}
\label{Dsol}
\begin{align}
\tilde\alpha_{ttxx}^{\text{(D)}} & =  -2 \partial_t  \partial_x (\varphi_t \varphi_x),\\
\tilde\beta_{ttyy}^{\text{(D)}} & =-2 \partial_t \partial_y (\varphi_t \varphi_y), \\
\tilde\gamma^{\text{(D)}}_{xx yy} & =-2 \partial_x \partial_y(\varphi_x \varphi_y),
\end{align}
\end{subequations}
where the subscript (D) implies the use of diagonal equations. The latter result is to be compared to~(\ref{eq:constraint_alpha_3})-(\ref{eq:constraint_gamma_3}). It is not difficult to see that the expressions for $\alpha$, $\beta$ and $\gamma$ are equivalent up to extra derivatives in the case of diagonal equations. As we discussed above, however, within the wave turbulence regime, integration with respect to spatial coordinates or the time coordinate is equivalent to dividing by $ik$ or $-i\omega_k$. 
Therefore, under the assumption of the wave turbulent regime, the two sets of equations are not only compatible but, in fact, equivalent. This is also the reason why, in this regime, the integration "constants" $c_0$, $c_1$ and $c_2$ in~(\ref{Bianchi_sol}) are equal to $0$. We also provide an alternative derivation of the equivalence between the two sets of equations, based on a Fourier decomposition in appendix~\ref{app:equivalence_wave_turbulence}.

Consequently, within the weak wave–turbulence regime, one may consistently choose to solve either the off-diagonal or the diagonal equations. It should be noted, however, that this equivalence does not hold in a more general setting (strong gravitational wave turbulence), where discrepancies may arise at fourth order in the perturbative expansion.

\subsection{Statistical predictions}

In \cite{galtier_2017} and \cite{gay_2024}, the off-diagonal equations were used to obtain results in the weak turbulence regime. They lead to the following expressions:
\begin{subequations}
\begin{align}
    \label{eq:alpha}
    \tilde{\alpha}(\mathbf{k}) &=
	- \frac{1}{2} \int_{\mathbb{R}^4} \sum_{s_1, s_2} 
	\frac{s_1 k_1 p_2 + s_2 k_2 p_1}{p \sqrt{k_1 k_2}}
	\frac{a^{s_1}_{\boldsymbol{k}_1} a^{s_2}_{\boldsymbol{k}_2}}{s_1 k_1 + s_2 k_2} \\\notag
    & \hspace{85pt} \times
    e^{i \Omega_{1 2} t} \delta^0_{12}(\boldsymbol{k})
	\mathrm{d}^2\boldsymbol{k}_1 \mathrm{d}^2\boldsymbol{k}_2 \\
	\label{eq:beta}
	\tilde{\beta}(\mathbf{k}) &=
	- \frac{1}{2} \int_{\mathbb{R}^4} \sum_{s_1, s_2} 
	\frac{s_1 k_1 q_2 + s_2 k_2 q_1}{q \sqrt{k_1 k_2}}
	\frac{a^{s_1}_{\boldsymbol{k}_1} a^{s_2}_{\boldsymbol{k}_2}}{s_1 k_1 + s_2 k_2} \\\notag
    & \hspace{85pt} \times
    e^{i \Omega_{1 2} t} \delta^0_{12}(\boldsymbol{k})
	\mathrm{d}^2\boldsymbol{k}_1 \mathrm{d}^2\boldsymbol{k}_2, \\
    \label{eq:gamma}
	\tilde{\gamma}(\mathbf{k}) &=
	- \frac{1}{2} \int_{\mathbb{R}^4} \sum_{s_1, s_2} 
	\frac{p_1 q_2 + p_2 q_1}{pq \sqrt{k_1 k_2}}
	a^{s_1}_{\boldsymbol{k}_1} a^{s_2}_{\boldsymbol{k}_2} \\\notag
    & \hspace{85pt} \times
    e^{i \Omega_{1 2} t} \delta^0_{12}(\boldsymbol{k})
	\mathrm{d}^2\boldsymbol{k}_1 \mathrm{d}^2\boldsymbol{k}_2,
\end{align}
\end{subequations}
where we used the shorthand notations $\delta^0_{12}(\boldsymbol{k}) = \delta (\boldsymbol{k} - \boldsymbol{k}_1 - \boldsymbol{k}_2)$ and $\Omega^{0}_{12} = s k - s_1 k_1 - s_2 k_2$.

These decompositions are then substituted into the equation~(\ref{eq:dynamic_3}), reducing the system to a single integro-differential equation governing the canonical variables $a^s_{\boldsymbol{k}}$, commonly referred to as the amplitude equation. It takes the form:
\begin{equation}
    \partial_t a^{s}_{\boldsymbol{k}} = 
    \epsilon^2
    \int_{\mathbb{R}^6} 
    \mathcal{T}^{s s_1 s_2 s_3}_{\boldsymbol{k} \boldsymbol{k}_1 \boldsymbol{k}_2 \boldsymbol{k}_3}
    a^{s_1}_{\boldsymbol{k}_1} a^{s_2}_{\boldsymbol{k}_2} a^{s_3}_{\boldsymbol{k}_3} \\[10pt]
    e^{i\Omega^{0}_{123} t} \delta^{0}_{123}(\boldsymbol{k})
    \prod_{i = 1}^3
    \mathrm{d}^2\boldsymbol{k}_i,
\label{eq:amplitude_equation}
\end{equation}
where $\Omega^{0}_{123} = s k - s_1 k_1 - s_2 k_2 - s_3 k_3$, and
$\mathcal{T}^{s s_1 s_2 s_3}_{\boldsymbol{k} \boldsymbol{k}_1 \boldsymbol{k}_2 \boldsymbol{k}_3}$
denotes the interaction coefficient. Its explicit expression is not reproduced here and may be found in \cite{galtier_2017, gay_2024}.

The amplitude equation \eqref{eq:amplitude_equation} can then be used to derive the statistical properties of the turbulent regime. This is achieved by applying a multiple time-scale expansion \cite{gay_2024}, yielding an evolution equation for the wave-action spectrum. Assuming statistically homogeneous turbulence, one defines
\begin{equation}
\begin{aligned}
    n_{\boldsymbol{k}} (t) 
    &= \sum_{s = \pm} \left\langle a^s_{\boldsymbol{k}} (t) \, a^{-s}_{-\boldsymbol{k}} (t) \right\rangle \\
    &= \left\langle
        k \left| \hat \phi_{\boldsymbol{k}}(t) \right|^2
        + \frac{1}{k} \left| \partial_t \hat \phi_{\boldsymbol{k}}(t) \right|^2
    \right\rangle ,
\end{aligned}
\end{equation}
where $\langle \cdot \rangle$ denotes an ensemble average.
This quantity, which is related to the energy spectrum through $e(\boldsymbol{k}) = k \, n(\boldsymbol{k})$, may be interpreted as the number of waves at wave vector $\boldsymbol{k}$.
As shown in \cite{galtier_2017, gay_2024}, wave action (N) and energy (E) defined by:
\begin{equation}
    N = \int_{\mathbb{R}^2} n_{\boldsymbol{k}}~\mathrm{d}\boldsymbol{k}
    \quad
    \text{and}
    \quad
    E = \int_{\mathbb{R}^2} e_{\boldsymbol{k}}~\mathrm{d}\boldsymbol{k}
\end{equation} are two conserved quantities. It should be noted that, while energy is always conserved, wave action requires the existence of specific symmetries, which are satisfied in this case.

The time evolution of $n_{\boldsymbol{k}}$ is governed by the kinetic equation
\begin{multline}
    \partial_t n_{\boldsymbol{k}} = 
    36 \pi \epsilon^4 
    \int_{\mathbb{R}^6} 
    \mathcal{L}^{\boldsymbol{k} \boldsymbol{k}_1}_{\boldsymbol{k}_2 \boldsymbol{k}_3}
    \left(
        \frac{1}{n_{\boldsymbol{k}}}
        + \frac{1}{n_{\boldsymbol{k}_1}}
        - \frac{1}{n_{\boldsymbol{k}_2}}
        - \frac{1}{n_{\boldsymbol{k}_3}}
    \right) \\[10pt]
    \times
    n_{\boldsymbol{k}}
    n_{\boldsymbol{k}_1}
    n_{\boldsymbol{k}_2}
    n_{\boldsymbol{k}_3}
    \,
    \delta^{01}_{23}(k)
    \,
    \delta^{01}_{23}(\boldsymbol{k})
    \,
    \mathrm{d}^2\boldsymbol{k}_{1}
    \mathrm{d}^2\boldsymbol{k}_{2}
    \mathrm{d}^2\boldsymbol{k}_{3} ,
\label{eq:kinetic}
\end{multline}
where $\mathcal{L}^{\boldsymbol{k} \boldsymbol{k}_1}_{\boldsymbol{k}_2 \boldsymbol{k}_3}$ denotes the collision kernel, whose explicit expression is not reproduced here (see \cite{galtier_2017, gay_2024}). The Dirac delta functions
$\delta^{01}_{23}(k) = \delta \left( k + k_1 - k_2 - k_3 \right)$
and
$\delta^{01}_{23}(\boldsymbol{k}) = \delta \!\left( \boldsymbol{k} + \boldsymbol{k}_1 - \boldsymbol{k}_2 - \boldsymbol{k}_3 \right)$
define the resonant manifold on which non-linear exchanges of wave action and energy occur.

By performing the angular integration and assuming scale invariance of the one-dimensional wave-action spectrum,
\begin{equation}
    N(k) = \int_{0}^{2\pi} k \, n_{\boldsymbol{k}} \, \mathrm{d}\theta \propto k^{x},
\end{equation}
the kinetic equation \eqref{eq:kinetic} admits four power-law solutions. Two of these, corresponding to $x = 0$ and $x = 1$, describe stationary regimes with vanishing fluxes and are associated with equipartition of wave action and energy, respectively. 
The remaining two solutions, $x = -2/3$ and $x = -1$, are called the Kolmogorov-Zakharov spectra. They are of greater physical relevance: the former corresponds to an inverse wave-action cascade, while the latter describes a direct energy cascade. 

To our knowledge, these spectra are the first exact statistical solutions ever obtained in gravitational wave turbulence. 

\section{numerical code (TIGER)}

This section presents an overview of the numerical  \texttt{TIGeR} code (\textbf{T}urbulence \textbf{I}n \textbf{Ge}neral \textbf{R}elativity),  employed to solve the system of equations given by the equations (\ref{eq:constraint_alpha_3}--\ref{eq:constraint_gamma_3}) and \eqref{eq:dynamic_3}.


This simulation is a Direct Numerical Simulation code using the Python library CuPy \cite{cupy_learningsys2017}. Spatial integration is performed using a pseudo-spectral method \cite{orszag_1969}, while time integration employs an explicit Adams-Bashforth method of order two \cite{demailly_2006} with a fixed time step $dt$.

The governing equations are solved on a square domain of size $L \times L$ discretised using a numerical grid of resolution $N \times N$. To maintain numerical stability and accuracy, dealiasing is implemented using the standard $2/3$-rule, and a dissipative term is introduced at small scales to prevent the accumulation of energy in high-frequency modes.

As in \cite{galtier_2021}, the initial condition consists of a narrow-band distribution in Fourier space, modulated by random phases. This condition is imposed on the normal variables:
\begin{equation}
    \phi_{\boldsymbol{k}}(t=0) = 
    A_i 
    \left[ \boldsymbol{k}^2 - (k_i - 1)^2 \right]
    \left[ (k_i + 1)^2 - \boldsymbol{k}^2 \right]
    e^{2i \pi \psi_{\boldsymbol{k}}},
\end{equation}
where $k_i$ is the injection wave-number and $\boldsymbol{k} \in \left[(k_i -1) / \sqrt{2}, (k_i + 1) / \sqrt{2} \right]^2$. The coefficient $A_i$ represents the amplitude of the initial perturbation. It is chosen such that the total wave action initially injected into the system equals $n_i$, and $\psi_{\boldsymbol{k}}$ is a real-valued random variable uniformly distributed in the interval $\left[0, 1 \right[$. 

From this spectral initial condition, we reconstruct the corresponding physical-space field $\phi$ through an inverse Fourier transform. Then, we reconstruct the $\gamma$ field by using the constraint \eqref{eq:constraint_gamma_3} and $\alpha$ and $\beta$ are initialised using the equation (\ref{eq:G_zz_3}) by imposing that initially:
\begin{equation}
    \alpha_{xx} =  \beta_{yy} = - \frac{1}{2} \left(  \phi_x^2 + \phi_y^2 \right).
\end{equation}

In order to stabilise high-frequency modes, a dissipating function  is introduced, defined as:
\begin{equation}
    \mathcal{D}_{\boldsymbol{k}} =
    \begin{cases}
        - \nu_h k^4 & \mathrm{if}~k > k_h, \\
        0 & \mathrm{else},
    \end{cases}
\end{equation}
where $\nu_h$ is a dissipation coefficient, and $k_h$ is the wave number at which dissipation occurs. The dissipation is applied on the $\phi$ and $\phi_t$ fields, which is equivalent to using the dissipation directly upon the normal variables. Given that the dynamics of $\tilde \alpha$, $\tilde \beta$, and $\tilde \gamma$ are constrained by those of $\phi$ and $\phi_t$, it is unnecessary to apply dissipation to these variables. Moreover, no large-scale dissipation is imposed, as the energy cascade does not develop significantly towards the infrared region within the timescales considered.

Accordingly, at each time step, a Strang splitting method \cite{strang_1968} is employed: the fields are first evolved according to the system of equations \eqref{eq:main_system}, after which dissipation is applied via exponential filtering, which is equivalent to a Crank–Nicolson scheme.

All simulations are performed on an NVIDIA A100 Tensor Core GPU. This combination of hardware and optimised software accelerates code execution by up to a factor of 200 compared with the original implementation in \cite{galtier_2021}. The principal simulation parameters are summarised in Table~\ref{tab:simulation_parameters}.

\begin{table*}
    \centering
    \begin{tabular}{cccccccc}
        \toprule
        $N$ & $L$ & $dt$ & $k_i / \mathrm{d}k$ & $n_i$ & $k_h / \mathrm{d}k$ & $\nu_h$ & $t_f / t_\mathrm{GW}$ \\ 
        \midrule
        $1024$ & $\pi / 2$ & $5 \cdot 10^{-6}$ & $50$ & $200$ & $340$ & $1 \cdot 10^{-11}$ & $90 \cdot 10^{3}$ \\
        \bottomrule
    \end{tabular}
    \caption{Summary of the different parameters used in the simulations. $N$ is its resolution along one axis, $L$ is the size of the physical square box, $dt$ is the time step of the Adams-Bashforth integrator, $k_i$ is the injection wave number (normalised by $\mathrm{d}k = 2 \pi / L$ in order to depend on the resolution solely), $n_i$ is the total amount of wave action. Dissipation is controlled by $k_h$ and $\nu_h$, which respectively refer to the wavenumber at which dissipation occurs and the dissipation coefficient. The final time $t_f$ is normalized by the linear time $t_\mathrm{GW} := 1 / k_i$.}
    \label{tab:simulation_parameters}
\end{table*}
\section{Towards a direct numerical simulation of gravitational wave turbulence}

In this section, we present numerical results obtained with the TIGeR code. As we will see, the results obtained are consistent with several theoretical predictions regarding gravitational wave turbulence. However, the study also reveals the difficulty of simultaneously satisfying all of Einstein's equations with the required level of accuracy. As a result, the simulated system does not strictly correspond to general relativity, but should rather be regarded as an approximation to Einstein’s equations.

\subsection{Conservation of invariants and dual cascade}

\begin{figure}[hbtp]
    \centering
    \includegraphics[width=0.95\linewidth]{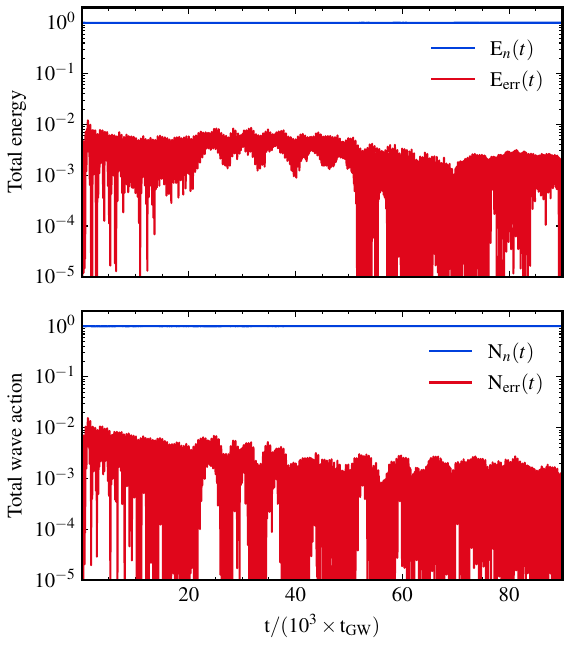}
    \caption{Temporal evolution of the conserved quantities $E$ (top) and $N$ (bottom). In both plots, the blue line ($E_\mathrm{n}$ and $N_\mathrm{n}$) represents the value normalised by its initial condition, while the red line shows the corresponding relative error ($E_\mathrm{err}$ and $N_\mathrm{err}$).}
    \label{fig:etnt}
\end{figure}

We begin by examining the evolution of the two conserved quantities: total wave action $N = \int_k N(k)~\mathrm{d}k$, and total energy $E = \int_k k N(k)~\mathrm{d}k$ \cite{galtier_2017, gay_2024}. Their time dependence is shown in Figure~\ref{fig:etnt}. Although both quantities are theoretically conserved, small oscillations are observed, with relative amplitudes remaining mainly below $10^{-2}$.

\begin{figure*}[hbtp]
    \includegraphics[width=0.475\linewidth]{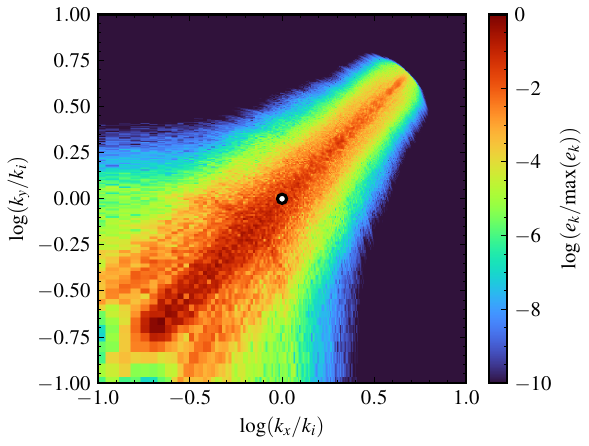}
    \includegraphics[width=0.475\linewidth]{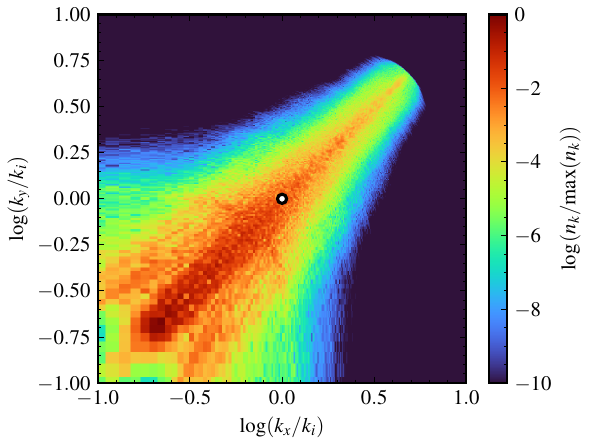}
    \caption{Energy (left) and wave action (right) two-dimensional spectra at time $t = 80 \times 10^3~t_\mathrm {GW}$ in a logarithmic grid. 
    The injection wavenumber normalises the axes, and a small white disk localises the initial excitation of the spectra.}
    \label{fig:2d_spectra}
\end{figure*}

\begin{figure*}[hbtp]
    \includegraphics[width=0.475\linewidth]{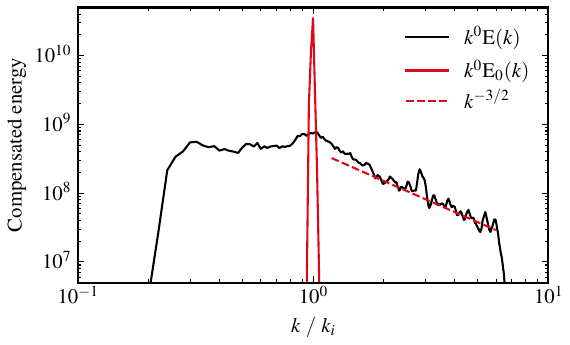}
    \includegraphics[width=0.475\linewidth]{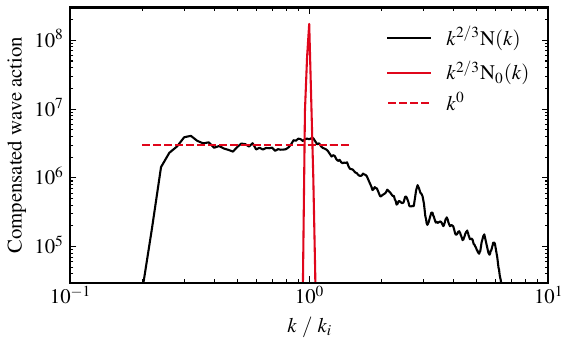}
    \caption{Energy (left) and wave action (right) one-dimensional spectra at times $t = 0$ (red) and $t = 80 \times 10^3~t_\mathrm {GW}$ (black) compensated by the Kolmogorov-Zakharov spectra $k^{0}$ and $k^{-2/3}$, respectively. 
    An empirical observation of the spectrum is given by a red dashed line in the two panels. The injection wavenumber normalises the horizontal axis.}
    \label{fig:1d_spectra}
\end{figure*}

As noted in \cite{galtier_2021}, once the front of the direct cascade reaches the dissipative range (from $t \sim 55\cdot 10^3 t_\mathrm{GW}$), the energy and wave action are expected to begin decaying according to a specific power law. However, such behaviour is not observed in the present simulation. To understand this discrepancy, we need to examine the figures \ref{fig:2d_spectra} and \ref{fig:1d_spectra}, which present the energy and wave-action spectra in their two-dimensional and one-dimensional representations, respectively. As seen in the plots, only a small fraction of the total energy and wave action is dissipated at small scales, owing to the very slow progression of the cascade in that direction.

The two-dimensional spectra in Figure \ref{fig:2d_spectra} provide valuable insight into the range of modes excited by the cascade dynamics. Note that the circular arc observed at $k \sim 320 \times 2\pi / L$ reflects the influence of dissipation at small scales. Nonetheless, the cascade spans a broad region of Fourier space, allowing for meaningful statistical analysis of the turbulent state. 

Both energy and wave action reach their maximum at low wavenumbers, indicating that an inverse cascade predominantly governs the system. As explained above, this inverse cascade is driven by wave action for which the flux is negative. However, since energy is proportional to it, it also propagates at small $k$. At small wavenumbers, the two-dimensional spectra exhibit slight angular anisotropy, suggesting that the cascades do not fully isotropise. In contrast, the spectral structure at high wavenumbers appears more focused, implying that the inverse cascade spreads more effectively in angle than the direct cascade. Finally, the energy distribution in the region of direct cascade is slightly more homogeneous than that of the wave action. 

The corresponding one-dimensional spectra, shown in Figure \ref{fig:1d_spectra}, are computed by averaging over five spectra, each separated by $\Delta t = 250 t_\mathrm{GW}$, and then applying a sliding average over 5 points. This procedure improves the statistical reliability of the spectra by suppressing numerical artefacts. The resulting spectrum for the inverse cascade is in good agreement with the theoretical prediction, with clear scaling in $k^{-2/3}$. However, the direct energy cascade follows a $k^{-3/2}$ scaling significantly different from the theoretical prediction in $k^{0}$. This property could be a signature of the domination of non-local interactions at small scales. 

\begin{figure}[ht]
    \centering
    \includegraphics[width=0.95\linewidth]{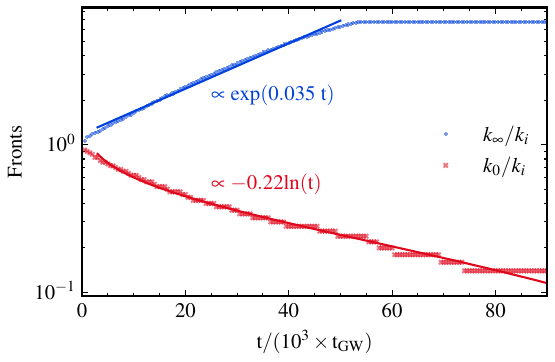}
    \caption{Temporal evolution of the fronts (in units of $k_i$) for the direct (blue) and inverse (red) cascades. 
    A linear regression is used to estimate their variations (solid lines).}
    \label{fig:fronts}
\end{figure}

\begin{figure*}[hbtp]
    \centering
    \includegraphics[width=0.95\linewidth]{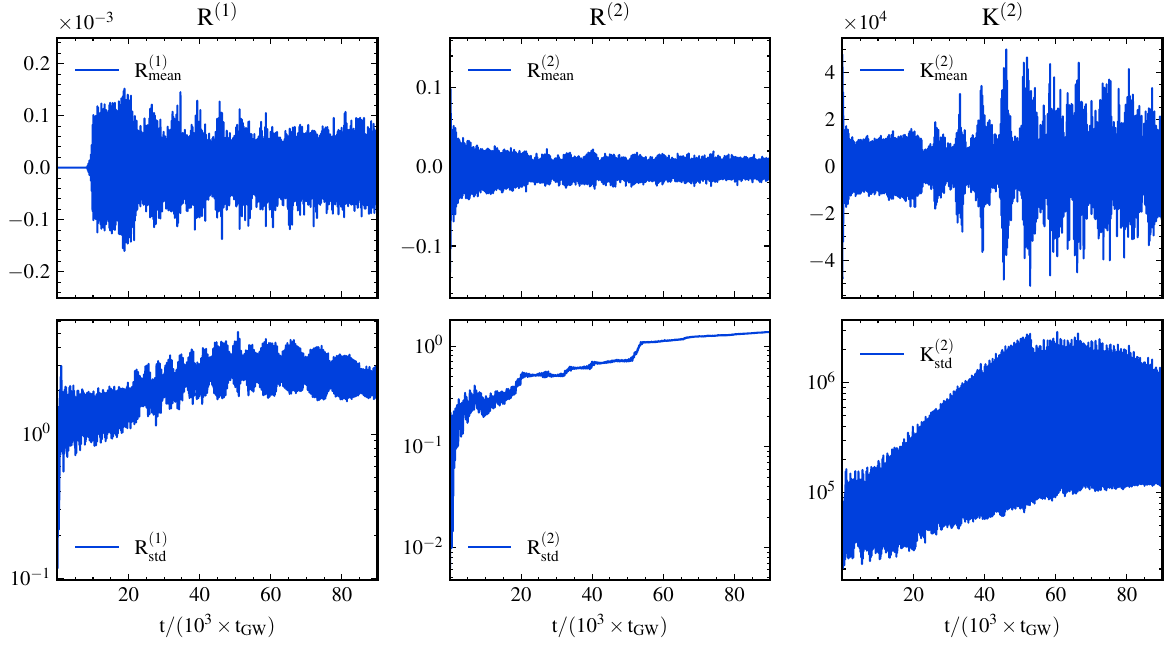}
    \caption{Time evolution of the mean (top) and the standard deviation (bottom) values of the Ricci and Krestchmann scalars.}
    \label{fig:scalars_stats}
\end{figure*}

Figure~\ref{fig:fronts} shows the temporal evolution of the front positions for both the direct and inverse cascades over the course of the simulation. The plateau reached by the direct cascade front marks the onset of dissipation, identifying the timescale at which the cascade ceases. Notably, our findings diverge from those presented in the subsequent section (and in \cite{gay_2025}), where the inverse cascade is observed to propagate considerably more rapidly than the direct cascade. Although this deviation may suggest a limitation of the super-local quartic-interaction model for gravitational-wave turbulence, another plausible explanation for this discrepancy lies in the relative sizes of the inertial zones: those obtained in direct numerical simulations are much smaller than those accessible in non-linear diffusion models. Consequently, the cascades do not extend over a sufficiently broad range to accentuate the difference in their propagation speeds. It is also conceivable that the observed slowing of the cascade is a finite-size effect. As discussed in \cite{zakharov_2005a}, the number of available modes diminishes at larger scales, thereby reducing the number of resonant quartic interactions. 

\subsection{Curvature invariants}

Throughout the simulation, we compute and compare the dynamics of the Ricci scalar $R$ and the Kretschmann scalar $K$, to numerically demonstrate that wave turbulence corresponds to the propagation of a physical degree of freedom and is not a pure gauge effect. We remind that for a general spacetime metric $g_{\mu \nu}$, these two scalar invariants are expressed by:
\begin{subequations}
\begin{align}
    R &= g^{\mu \nu} R^\rho_{\phantom{a} \mu \rho \nu}, \\
    K &= R^{\rho \sigma \mu \nu} R_{\rho \sigma \mu \nu},
\end{align}
\end{subequations}
where $R^\rho_{\phantom{a} \mu \rho \nu}$ is the Riemann tensor. Their expressions in the Hadad-Zakharov metric can be computed explicitly, but since they are quite cumbersome, we do not reproduce them here. 

However, it is much more relevant to expand their expressions in powers of $\epsilon$ and to focus on the first two orders  (to be consistent with the second-order resolution of Einstein's equation) and to study the evolution of these terms. Hence, we have $R=R^{(0)} + \epsilon R^{(1)} + \epsilon^2 R^{(2)} + o(\epsilon^2)$ and a  direct calculation gives the following expressions for the first two terms,
\begin{subequations}
\allowdisplaybreaks
\begin{align}
    R^{(1)} &= 
    2 \left( 
         \phi_{tt} 
        -  \phi_{xx} 
        - \phi_{yy}
    \right), \\[5pt]
    R^{(2)} &= 
    2 \big[ 
    \phi_t ^2 
    -  \phi_x^2 
    - \phi_y^2
    - \phi R^{(1)} +  \left( \alpha_{tt} + \beta_{tt} \right)
    \\[5pt]\notag
    & \hspace{20pt}
    - \left( \alpha_{xx} + \gamma_{xx} \right) 
    -  \left( \beta_{yy} + \gamma_{yy} \right) 
    \big].
\end{align}
\end{subequations}
We proceed in the same way for  the Kretschmann, $K=K^{(0)} + \epsilon K^{(1)} + \epsilon^2 K^{(2)} + o(\epsilon^2)$, and we show,
\begin{subequations}
\begin{align}
    K^{(1)} &= 0, \\[5pt]
    K^{(2)} &= 
    12 \big(
        \phi_{tt}^2 
        +\phi_{xx}^2 
        + \phi_{yy}^2 
    \big)
    \\[5pt]\notag
    & \hspace{10pt}
    + 8 \big(
         \phi_{xx}  \phi_{yy} 
        -  \phi_{tt}  \phi_{xx}
        -  \phi_{tt} \phi_{yy}
    \big)
    \\[5pt]\notag
    & \hspace{10pt}
    + 16 \big(
         \phi_{xy}^2 
        -  \phi_{tx}^2 
        -  \phi_{ty}^2
    \big).
\end{align}
\end{subequations}
Obviously, both $R^{(0)}$ and $K^{(0)}$ vanish (as the space-time reduces to Minkowski). Note that, since we are considering the Einstein equations in vacuum, $R$ must cancel out. However, this cannot be exactly the case for the ``truncated'' Ricci scalars $R^{(1)}$ and $R^{(2)}$, even though we expect them to be small compared to unity.

\begin{figure*}[hbtp]
    \centering
    \includegraphics[width=0.95\linewidth]{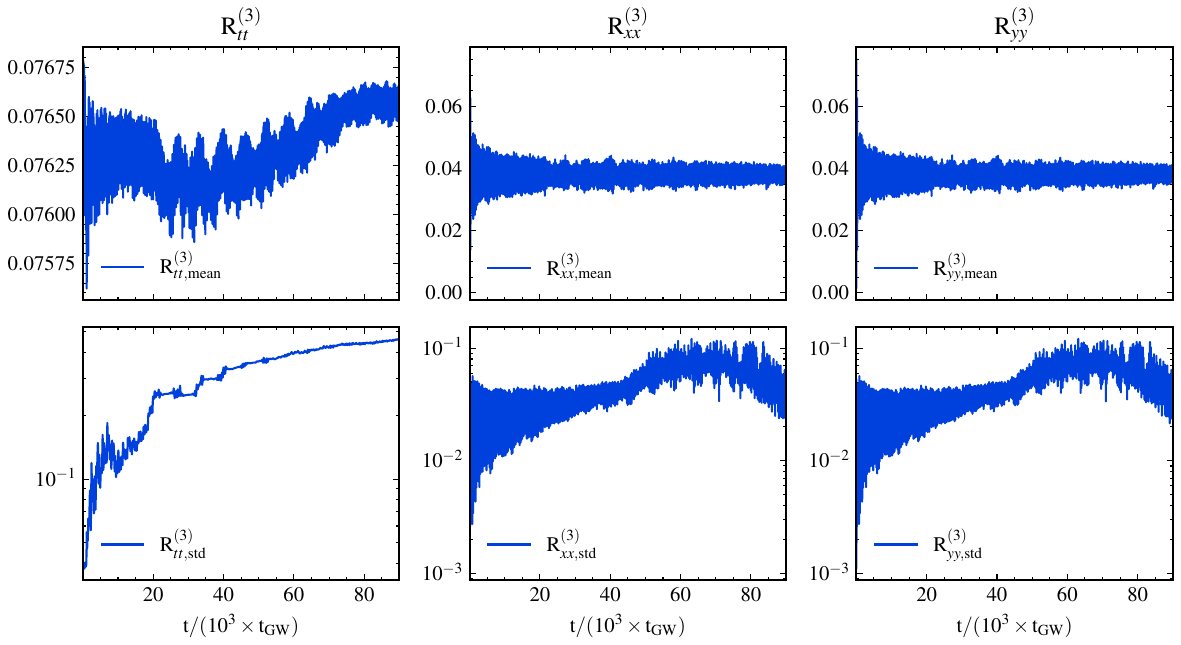}
    \caption{Time evolution of the mean (top) and the standard deviation (bottom) values of $\mathrm{R}_{zz}^{(3)}$ (equation \eqref{eq:G_zz_3}), $\mathrm{R}_{xx}^{(3)}$ (equation \eqref{eq:G_yy_3}) and $\mathrm{R}_{yy}^{(3)}$ (equation \eqref{eq:G_xx_3}).}
    \label{fig:constraints_stats}
\end{figure*}

Actually, $R^{(1)}$ serves as a diagnostic for the nonlinear nature of the system: deviations from zero indicate the presence of (weak) nonlinearities in the dynamics. On the other hand, $R^{(2)}$ is intrinsically linked to the constraint equations (\ref{eq:G_zz_3})--(\ref{eq:G_xx_3}), for it is a linear combination of them. Furthermore, $K^{(2)}$ provides a criterion for assessing the physical validity of the model, since the Kretschmann scalar must be non-zero due to the presence of gravitational waves, thus one may expect $\sqrt{|K|} \gg |R|$.

Figure~\ref{fig:scalars_stats} shows the temporal evolution of the spatial means and standard deviations (std) values of these scalar quantities. Due to the limitation of the pseudo-spectral resolution method, the spatial mean (corresponding to the mode $\boldsymbol{k} = (0, 0)$) is not properly resolved.
Hence, it provides insight into the precision of the solution, provided that the mean is smaller than the standard deviation. 

Observing the behaviour of these invariants allows us to identify three distinct regimes. First, the onset of oscillations in $R^{(1)}$ at $t \approx 10 \times 10^3 t_{GW}$ marks the emergence of nonlinear dynamics and is consistent with the theoretical prediction $t \sim t_\mathrm{GW} / \epsilon^4$, where $\epsilon \sim 10^{-1}$ denotes the amplitude of $\phi$.
Secondly, the slight change in the behaviour observed after $t > 20 \times 10^3 t_\mathrm{GW}$ coincides with the arrival of the inverse cascade front at modes $k_x = 0$ and $k_y = 0$. For these modes, the spatial integration of the constraint equations ensures that the fields are zero. This introduces a minor perturbation to the dynamics that seems to have no significant impact on the main conclusions of this study, as the essential physical processes occur near the $k_x = k_y$ mode line, where the dual cascade develops.
Finally, the alteration at $t \approx 55 \times 10^3$ coincides with the arrival of the direct front at the dissipative area.

We note that the mean is smaller than the standard deviation, especially for $R^{(1)}$ and $K^{(2)}$. The situation for $R^{(2)}$ can be clarified by examining the evolution of the diagonal equations, as represented in Figure~\ref{fig:constraints_stats}, which we will discuss further below. Additionally, it is important to highlight that the standard deviation of $R^{(1)}$ exceeds that of $R^{(2)}$ by a factor of three for $t > 55 \times 10^3 t_\mathrm{GW}$. Consequently, the magnitude of $R$ is effectively similar to that of $R^{(1)}$, which is on the order of unity. In contrast, the amplitude of $K$ is approximately equal (or larger) to that of $K^{(2)}$, making $ \sqrt{|K|}$ significantly greater than $R$. This observation confirms that the vacuum assumption is maintained throughout the simulation, thereby validating the physical propagation of a degree of freedom and ensuring the model's physical validity.

\begin{figure*}[hbtp]
    \centering
    \includegraphics[width=0.9\linewidth]{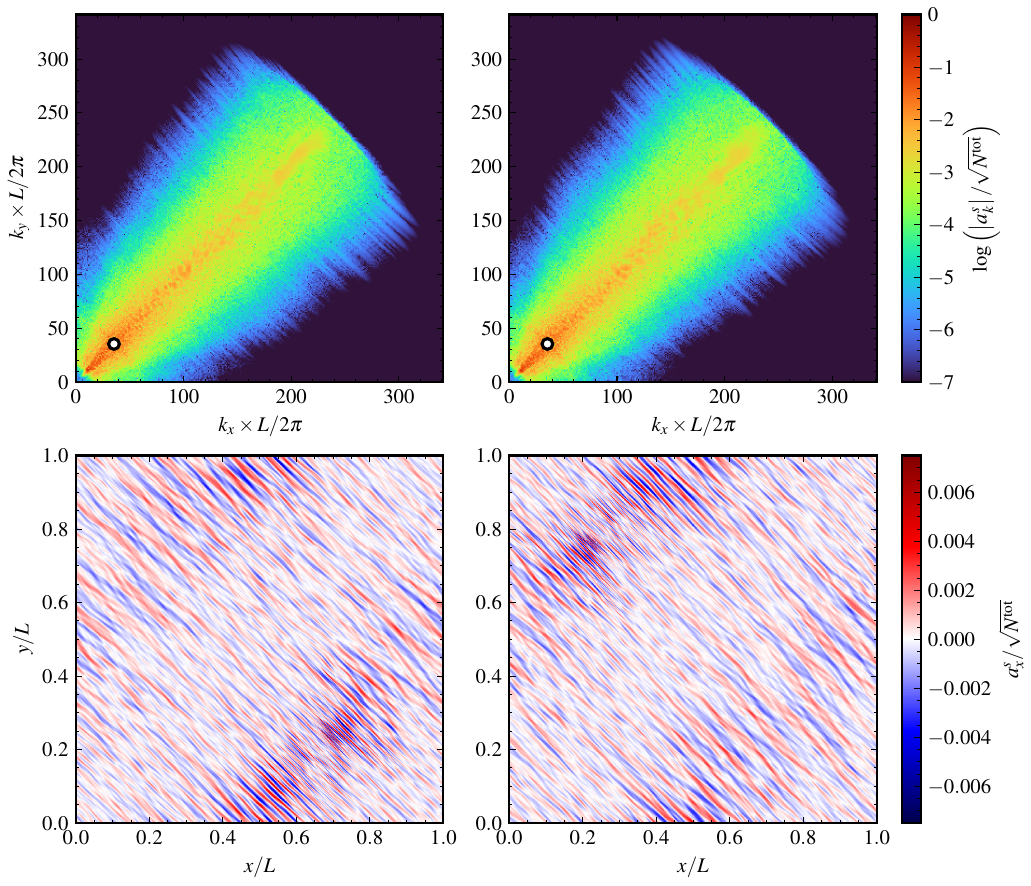}
    \caption{Canonical variables $a^+$ (left) and $a^-$ (right) in spectral (top) and physical (bottom) space at time $t = 80 \times 10^3~t_{\mathrm{GW}}$. The length of the box normalises the axes. In the spectral plots, white disks represent the initial excitation.}
    \label{fig:aks_spectra}
\end{figure*}

Figure~\ref{fig:constraints_stats} represents the time evolution of the diagonal equations (\ref{eq:G_zz_3})--(\ref{eq:G_xx_3}). As discussed earlier, the discrepancies should disappear if we solve equations (\ref{eq:constraint_alpha_3})--(\ref{eq:dynamic_3}) and if the wave turbulence regime holds true. However, our numerical results indicate that we still observe a non-zero mean field and a non-zero standard deviation. These errors can be attributed to artefacts of the pseudo-spectral resolution method. This method does not account for the dynamics of the mode with $k_x = 0$ or $k_y = 0$ (in particular, the mode $\boldsymbol{k} = 0$ corresponding to the mean). Specifically, relationships (\ref{eq:constraint_alpha_3})--(\ref{eq:constraint_gamma_3}) become incompatible when applying a pseudo-spectral technique, as the left-hand side has a zero mean while the right-hand side does not. The combination of this numerical integration error and finite-size effects (along with the need for dealiasing) results in a failure to satisfy the diagonal Einstein equations. Nevertheless, we note that these deviations are small, such that $|R| \ll \sqrt{|K|}$. Moreover, the statistical results of this simulation align well with wave turbulence predictions. Therefore, we will focus on the following aspects: probability density functions and structure functions.

\subsection{Statistical properties of the canonical variables}

We now turn our attention to the canonical variables (\ref{eq:canonical_variables_definition}), which are the natural fields of the wave-turbulence system. Figure~\ref{fig:aks_spectra} shows their variations in spectral and physical space at the end of the simulation $t = 80 \times 10^3 t_\mathrm{GW}$. It is observed that the two fields appear broadly similar across the two domains; however, in Fourier space, a small asymmetry is noted: $a_{\boldsymbol{k}}^+$ tends to develop more prominently above the line $k_x = k_y$, whereas $a_{\boldsymbol{k}}^-$ exhibits the opposite tendency. This slight discrepancy can be attributed to the random initial condition. Recall that the initial excitation is applied to $\phi_{\boldsymbol{k}}$ rather than $\partial_t \phi_{\boldsymbol{k}}$. According to expression~\eqref{eq:canonical_variables_definition}, this leads to the condition $a_{\boldsymbol{k}}^+ = a_{\boldsymbol{k}}^-$ at $t = 0$. Although this equality remains approximately valid at later times, it no longer holds exactly due to the emergence of non-linear effects and energy transfers resulting from wave–wave interactions.

\begin{figure*}[hbtp]
    \centering
    \includegraphics[width=0.85\linewidth]{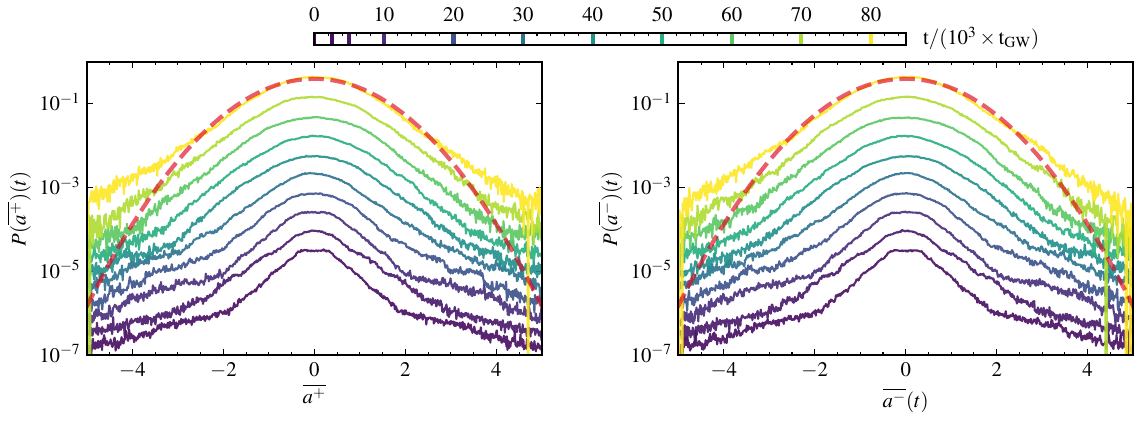}
    \caption{Normalized probability distribution function of canonical variables $\overline{a^+}(t)$ (left) and $\overline{a^-}(t)$ (right) at different simulation times. The curves have been vertically shifted by a factor of 2 to improve readability; higher curves correspond to later times. A standard Gaussian distribution is shown by a red dashed line and coincides with the probability distribution function at $t = 80 \times 10^3 t_\mathrm{GW}$.}
    \label{fig:aks_pdf_time}
\end{figure*}

\begin{figure*}[hbtp]
    \centering
    \includegraphics[width=0.85\linewidth]{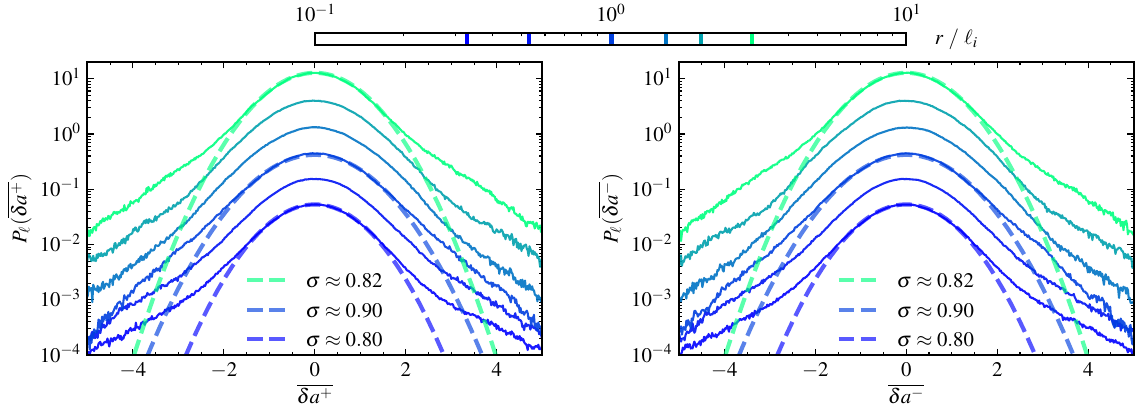}
    \caption{Normalized probability distribution function of the canonical variable increments $P_\ell(\overline{\delta a^+})$ (left) and $P_\ell(\overline{\delta a^-})$ (right) at $t = 80 \times 10^3~t_\mathrm{GW}$. The curves have been vertically shifted by a factor of 3 to improve readability; higher curves correspond to larger increments $\ell$. The centres of these probability distribution functions are compared with Gaussian distributions with a corresponding standard deviation determined by the behaviour of the PDF in the range $[-1, 1]$.}
    \label{fig:aks_pdf_space}
\end{figure*}

In the wave turbulence regime, wave amplitudes are assumed to be small to keep non-linearities weak. In such a situation, the two-point statistics in Fourier space remain close to Gaussianity, but not exactly Gaussian, to allow the regeneration of the cumulants via the product of lower-order cumulants  \citep{benney_1966,benney_1967c,galtier_2022}. The natural asymptotic closure relies precisely on this property. Under this condition, the probability distribution functions (PDF) of the canonical variables $a_{\boldsymbol{k}}^s$ are also expected to be close to Gaussianity. Since the Fourier transform preserves Gaussianity, we can also study the PDF in physical space and use the corresponding fields, $a^s(t, \boldsymbol{x})$, which are simpler to manipulate. 

We begin by tracking the evolution of the normalised probability distribution function of the canonical variables ($\overline{a^+}$ and $\overline{a^-}$) throughout the simulation. Here, the term "normalised" denotes that each probability distribution function is divided by its standard deviation, thereby enabling direct comparison with a standard Gaussian distribution. Note that all probability distribution functions discussed in this section have zero mean. The resulting distributions are shown in Figure \ref{fig:aks_pdf_time}. While the initial condition is clearly non-Gaussian, the probability distribution function progressively approaches a Gaussian form (over 2 decades in amplitude) as the simulation advances, consistent with the development of wave turbulence. However, the emergence of non-Gaussian tails is also observed, indicating an increasing likelihood of large-amplitude events. This behaviour suggests the formation and persistence of coherent structures throughout the simulation.

\begin{figure*}[ht]
    \centering
    \includegraphics[width=1.0\linewidth]{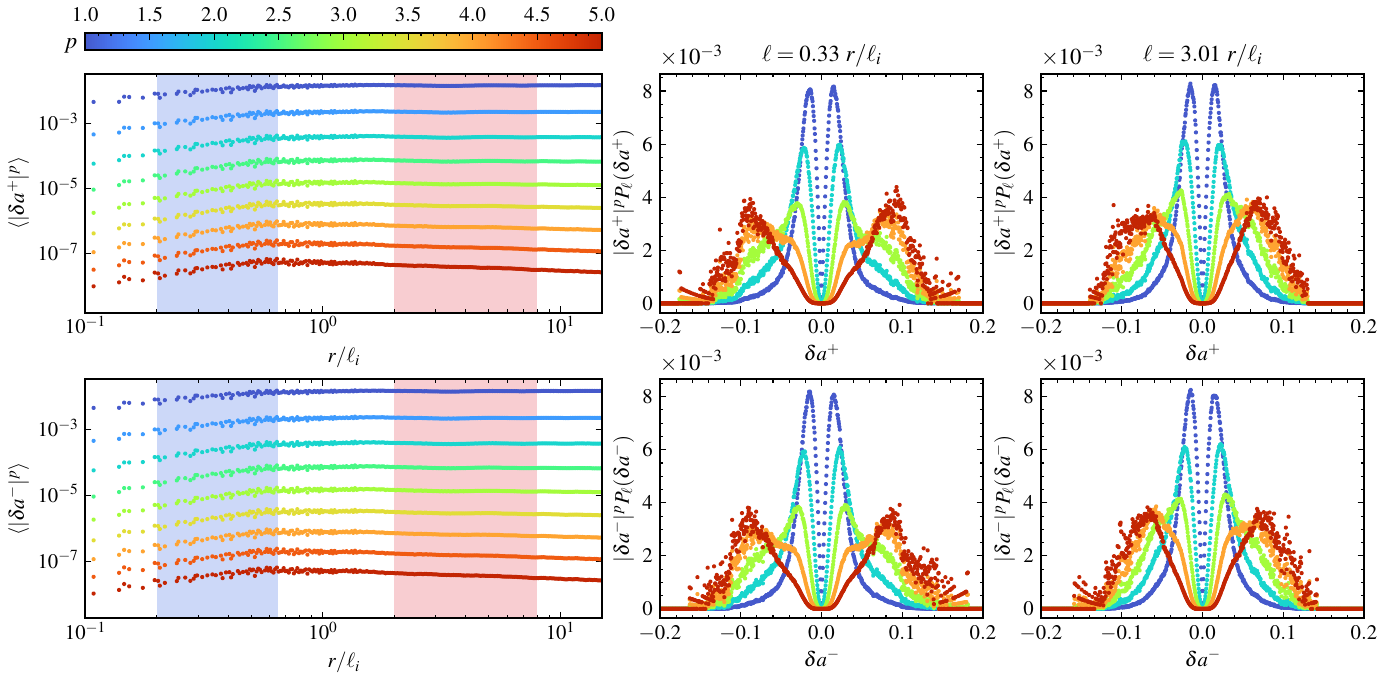}
    \caption{Structure functions for $a^+$ (top-left) and $a^-$ (bottom-left). The coloured area corresponds to the range of data used to compute the scaling exponents $\zeta(p)$. In the two columns on the right-hand side, we represent the corresponding integrands up to moment five for two different length scales within the inertial ranges. Statistical convergence begins to fail at the third moment.}
    \label{fig:aks_structures}
\end{figure*}

A second diagnostic involves analysing the probability distribution function of the canonical variable increments, defined as $\delta a^s = a^s(\boldsymbol{x} + \boldsymbol{r}) - a^s(\boldsymbol{x})$. In this study, the increment vector $\boldsymbol{r}$ is compared with the initial injection length scale, $\ell_i = 2\pi / k_i$. This quantity is useful for assessing the relative likelihood of turbulent fluctuations across different spatial scales. 
The corresponding results are presented in Figure \ref{fig:aks_pdf_space}, focusing on increments within the inertial ranges of the two cascades. The corresponding PDFs are compared with centred Gaussian probability distributions, plotted in dashed line, whose standard deviation is determined using the core of the PDF (values comprise between $[-1, 1]$). 
While the central parts of these distributions are clearly Gaussian, the tails exhibit significant deviations—clear signatures of non-Gaussian behaviour. Note that the probability distribution functions are computed by averaging over five outputs, each separated by $\Delta t = 250 t_\mathrm{GW}$. 
As the increment length approaches the size of the inverse-cascade condensate, the probability distribution function departs more strongly from Gaussianity, exhibiting broader tails and a more pronounced central peak. These features can be attributed to long-range correlations introduced by the condensate. Similarly, for increments of the order of the injection scale, the central region of the probability distribution function also deviates from Gaussianity, due to residual correlations from the initial condition that the turbulent dynamics have not fully dissipated.
These results demonstrate that the non-linear dynamics progressively smooth the initially random conditions and that the statistical properties of the canonical variables remain consistent with a system governed by weak wave turbulence. In a mere general wave-turbulent context, the results presented here are consistent with experimental data \cite{falcon_2007} or numerical simulation \cite{korotkevich_2008}.

The final diagnostic relies on the analysis of different order structure functions, with the $p^{\text{th}}$-order structure function: $S_p^s = \langle | \delta a^s |^p \rangle$ whose scaling law, in the inertial zone is: $\propto r^{\zeta^s(p)}$, where $\zeta^s(p)$ denotes the scaling exponent. 
\begin{figure}[ht]
    \centering
    \includegraphics[width=0.9\linewidth]{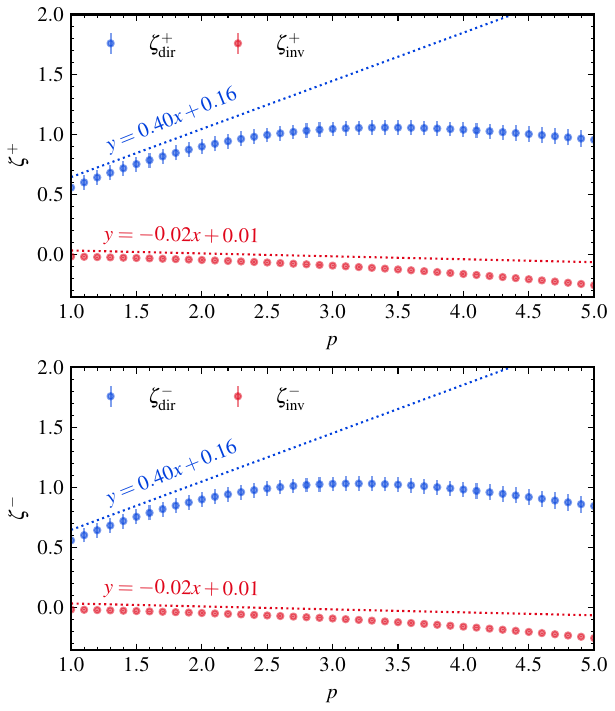}
    \caption{Scaling exponents of the canonical variables $a^+$ (top) and $a^-$ (bottom) in both the inverse inertial range (red) and the direct inertial range (blue).}
    \label{fig:aks_exponents}
\end{figure}
We estimate the structure functions for $p \in \left[1, 5 \right]$ by evaluating all possible translations on an $N \times N$ grid. The results are averaged by grouping values into prescribed bins. Specifically, we employ $400$ bins distributed over a geometrically spaced range of lengths, starting from 1 pixel (corresponding to a physical scale of $L / N$) up to $N / \sqrt{2}$ (i.e., $L / \sqrt{2}$). Note that the maximum value of $p$ is determined following the approach of \cite{dudokdewit_2004}. However, this value can be verified by assessing the convergence of the moments in both the inverse and direct cascade inertial area. In the present case, convergence start to fail for $p \geq 3$, meaning that the results above these values are not definitive. The resulting structure functions are presented in Figure \ref{fig:aks_structures}, with all lengths normalised by the injection scale. This normalisation facilitates the identification of distinct inertial ranges associated with the inverse and direct cascades.

\begin{figure*}[hbtp]
    \centering
    \includegraphics[width=0.85\linewidth]{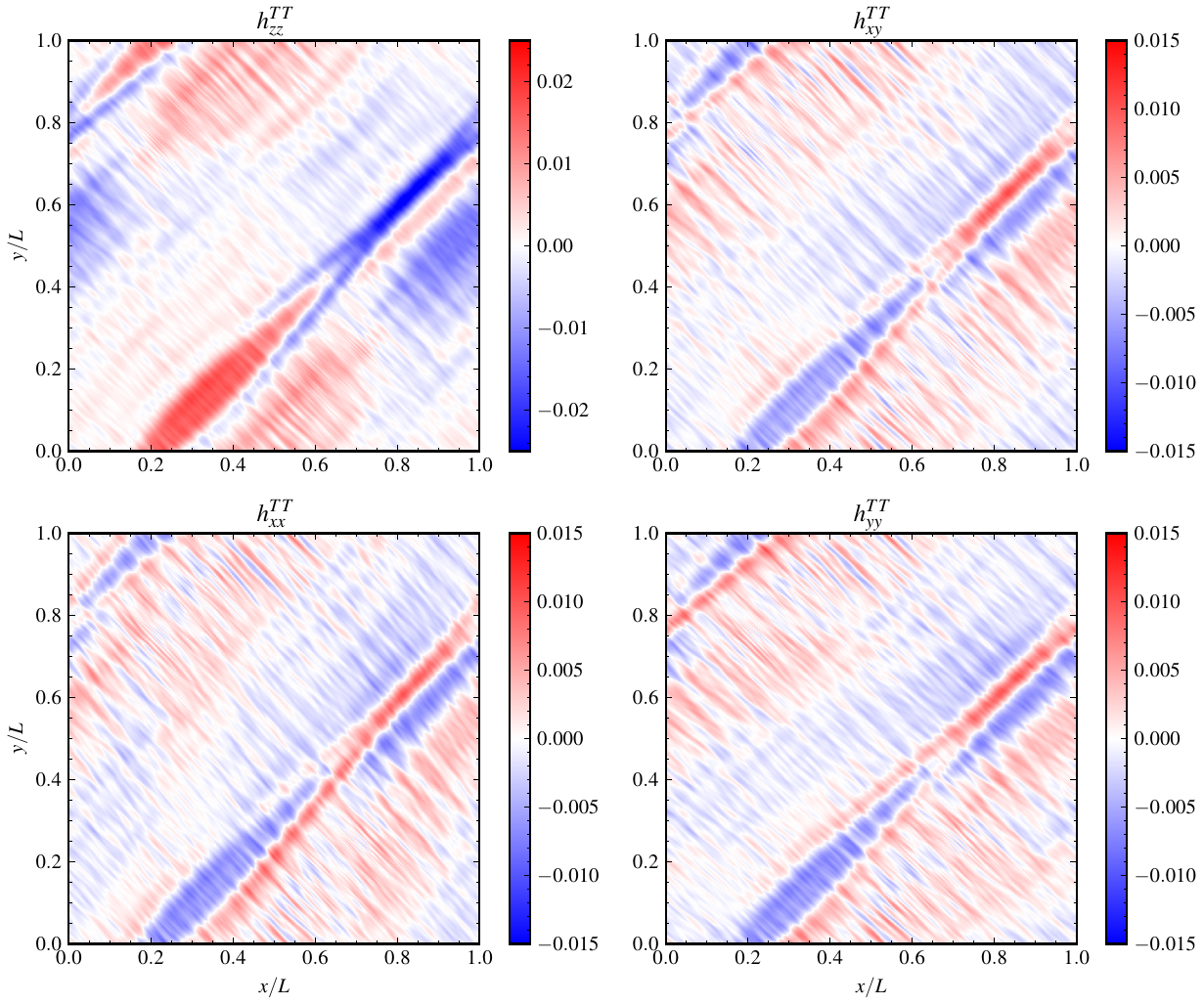}
    \caption{Gauge-invariant metric variables at $t = 80 \times 10^3 t_\mathrm{GW}$, $h^{TT}_{zz}$ (top left), $h^{TT}_{xy}$ (top right), $h^{TT}_{xx}$ (bottom left) and $h^{TT}_{yy}$ (bottom right).}
    \label{fig:metric}
\end{figure*}

Due to limited statistical sampling at small scales, the structure functions exhibit increased noise in this region. Nevertheless, within the identified inertial ranges, we estimate the scaling exponents by fitting the slopes and plot them as functions of $p$ in Figure~\ref{fig:aks_exponents}. The error, represented by vertical bars, is calculated from the linear regression residuals. This analysis provides insight into the fractal characteristics of gravitational wave turbulence. Although it is premature to draw definitive conclusions, it appears that the scaling laws for $\zeta^+$ and $\zeta^-$ are identical. Furthermore, there is no strong evidence of multi-fractal behaviour in the inverse-cascade regime. On the contrary, it appears to exhibit multifractal behaviour in the direct cascade regime, where the scaling exponent slope changes. These results demonstrate that the non-linear dynamics progressively smooth the initially random conditions. In a mere general wave-turbulent context, the results presented here are consistent with experimental data \cite{falcon_2007} or numerical simulation \cite{korotkevich_2008}.

As we will see below, the use of the gauge-invariant metric variables---i.e., physical observables in general relativity---leads to a more classical behaviour for the scaling exponents, characterised by a positive slope and monofractal behaviour.

\begin{figure*}[ht!]
    \centering
    \includegraphics[width=0.85\linewidth]{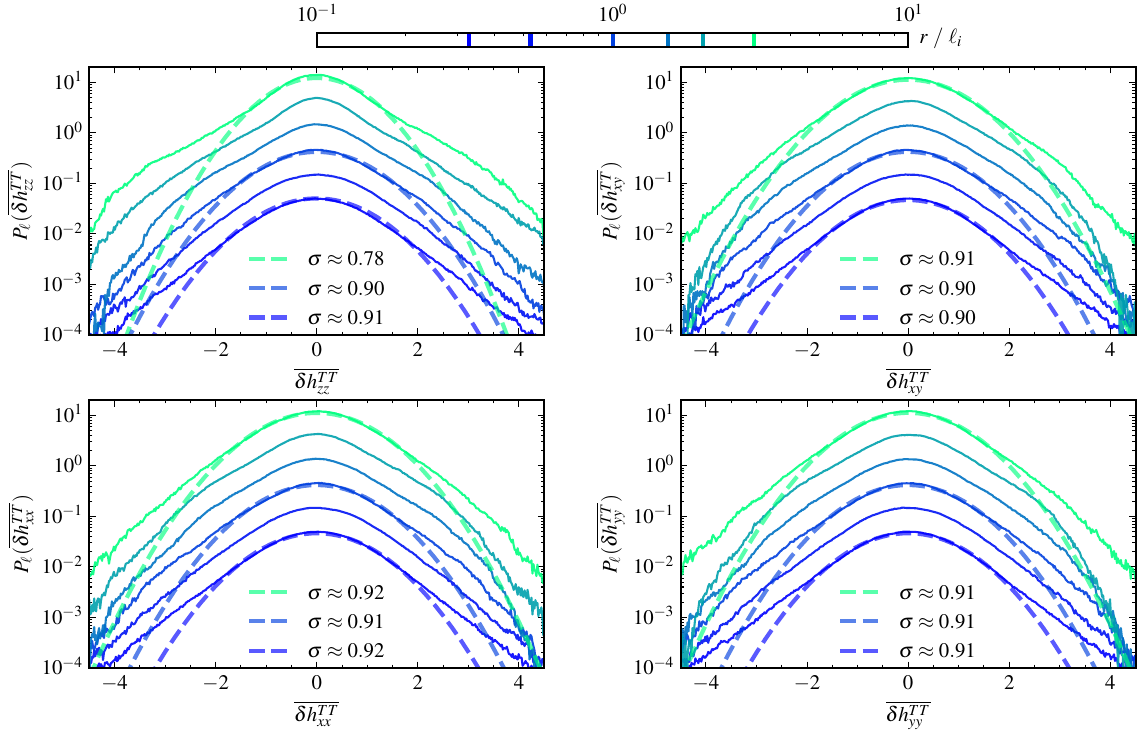}
    \caption{Normalized increment PDF of the gauge-invariant metric variables: $h^{TT}_{zz}$ (top left), $h^{TT}_{xy}$ (top right), $h^{TT}_{xx}$ (bottom left) and $h^{TT}_{yy}$ (bottom right). Curves are vertically shifted by a factor of 3 to improve readability: higher curves correspond to small increments. A standard canonical distribution function is shown as a dashed line, overlaid with the PDF increment at $r = \ell_i$.}
    \label{fig:metric_pdf}
\end{figure*}

\subsection{Statistical properties of the gauge-invariant metric variables}

We now proceed with the statistical analysis by examining the behaviour of the gauge-invariant metric variables (\ref{hTT}) $h^{TT}_{xx}$, $h^{TT}_{yy}$, $h^{TT}_{xy}$ and $h^{TT}_{zz}$, whose spatial profiles at the end of the simulation are displayed in Figure~\ref{fig:metric}. This examination provides insight into the relative amplitudes and dynamical properties of the individual components. As one can see, the 3 components of the transverse–traceless part of the metric, $h^{TT}_{xx}$, $h^{TT}_{yy}$ and $h^{TT}_{xy}$, exhibit comparable behaviour, with amplitudes spanning similar ranges. On the other hand, $h^{TT}_{zz}$ displays a slightly distinct pattern, characterised by a broader amplitude range. 

A further informative diagnostic is provided by the probability distribution functions of the field increments for the gauge-invariant variables, shown in Figure~\ref{fig:metric_pdf}. As previously, the increments are normalised by their respective standard deviations (all with zero mean) to enable a consistent comparison. The probability distribution functions are obtained by averaging over five outputs, each separated by $\Delta t = 250 t_\mathrm{GW}$. This analysis corroborates the earlier observation that the distributions of $h^{TT}_{zz}$ differ from those of $h^{TT}_{xx}$, $h^{TT}_{yy}$ and $h^{TT}_{xy}$. Specifically, the PDFs of the latter three components are closely similar, exhibiting Gaussian-like central regions together with non-Gaussian tails for both small and large increments. By contrast, $h^{TT}_{zz}$ shows a markedly different behaviour at large scales: its central region is distorted, and its tails display pronounced deviations from Gaussianity, indicating the presence of correlated high-amplitude events. The progressive transition towards a Gaussian-like distribution as the increment size increases is consistent with the interpretation that spatially distant events are more weakly correlated in amplitude.

\begin{figure*}[hbtp]
    \centering
    \includegraphics[width=1.\linewidth]{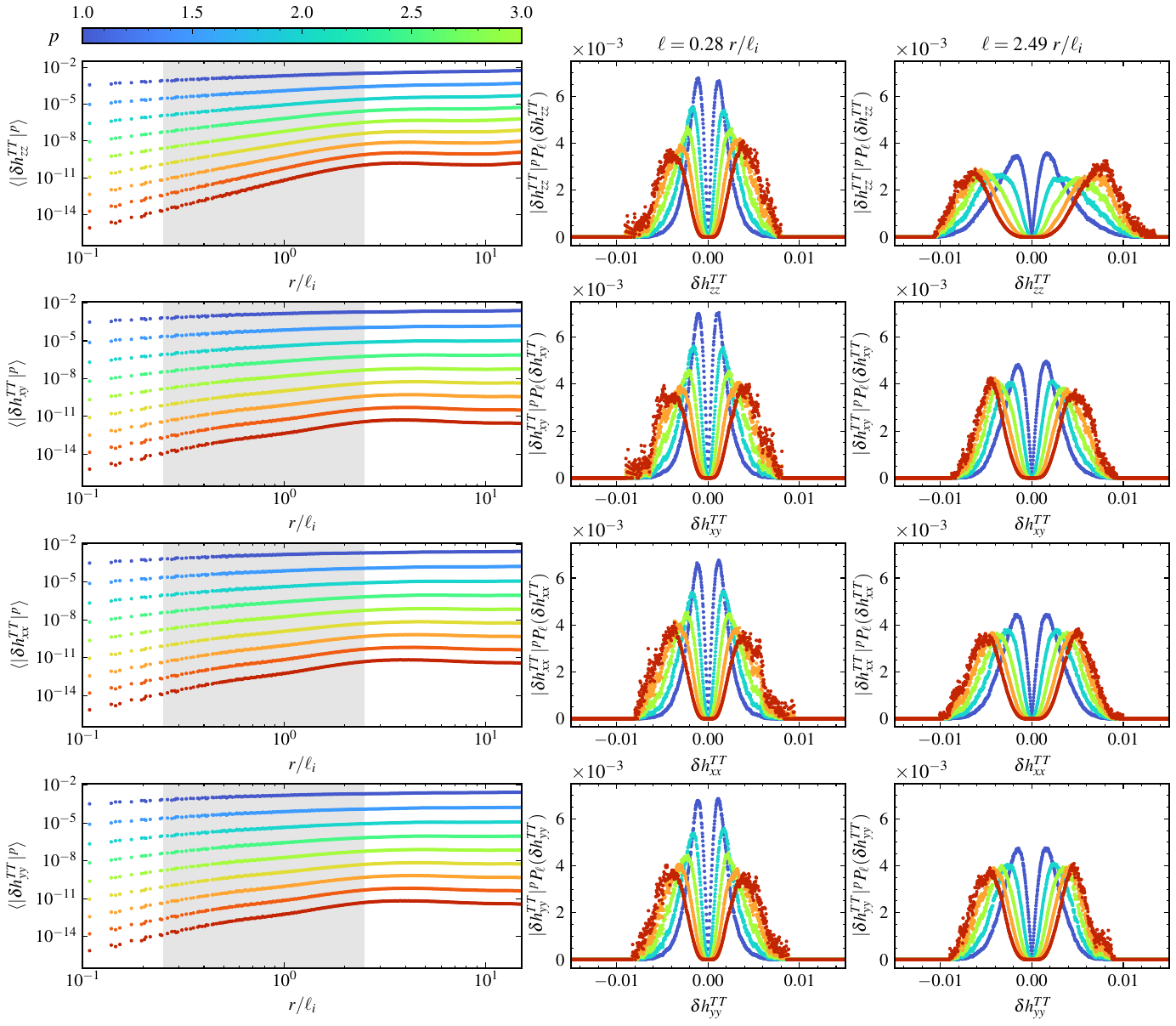}
    \caption{Structure functions for the gauge invariant metric variables. The grey area corresponds to the range of data used to compute the scaling exponents $\zeta(p)$. In the two columns on the right-hand side, we represent the corresponding integrands up to moment five for two different length scales at the edges of the inertial ranges. Statistical convergence begins to fail at the third moment.}
    \label{fig:metric_structures}
\end{figure*}

\begin{figure*}[ht!]
    \centering
    \includegraphics[width=0.85\linewidth]{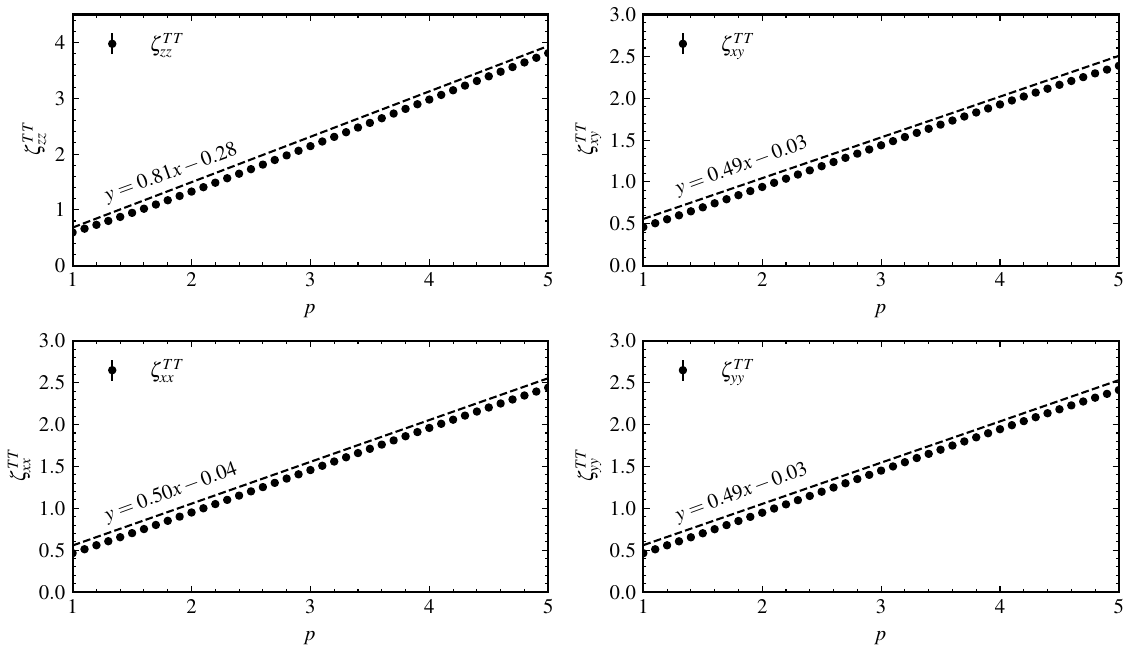}
    \caption{Scaling exponents of the gauge invariant metric.}
    \label{fig:g_exponents}
\end{figure*}

As in the case of the canonical variables, the evolution of the structure functions is presented in Figure~\ref{fig:metric_structures}, and the associated scaling exponents are shown in Figure~\ref{fig:g_exponents}. Notably, all gauge-invariant variables exhibit a single inertial range extending from $0.3$ to $3 r/\ell_i$. This interval is used to determine the corresponding scaling exponents, following the same procedure as described above: the exponents are obtained from linear fits to the slopes, plotted as functions of $p$, and the associated uncertainties are estimated from the residuals of the linear regressions. Here again, the maximum value of $p$ is determined following the approach of \cite{dudokdewit_2004} and is verified by assessing the convergence of the moments in the inertial area. In the present case, convergence starts to fail for $p \geq 3$, meaning that the results above these values are not definitive. 

Once again, we find that the behaviour of $h^{TT}_{xx}$, $h^{TT}_{yy}$ and $h^{TT}_{xy}$ differs from that of $h^{TT}_{zz}$. In particular, as $p$ increases, the structure function of $h^{TT}_{zz}$ becomes steeper than that of the other components, a result that is confirmed by the corresponding scaling exponents. All components display an approximately mono-fractal behaviour (the qualifier reflecting a slight but systematic deviation), characterised by linear scaling $y = 0.81 - 0.28$ for $h^{TT}_{zz}$ and $y = 0.50p - 0.03$ for the remaining components. Nevertheless, for all the components, a slight deviation from linearity is observed for $p \geq 3.0$, which is the starting point of the failure of convergence of the moments in the inertial area.
Note that such a mono-fractal behaviour is common in wave turbulence (see eg. \cite{vanbokhoven_2009,david_2024}), whereas strong turbulence is expected to be multi-fractal (see eg. \cite{anselmet_1984, frisch_95}). 
We can therefore conjecture that gauge-invariant metric variables represent more natural fields for studying gravitational wave turbulence in physical space.

\section{Conclusion}

The investigation presented here builds upon and extends the initial theoretical and numerical studies conducted in \cite{galtier_2017} and \cite{galtier_2021}, respectively. From a theoretical perspective, we provide a physical interpretation of the Hadad-Zakharov ansatz, treating it as 3D gravity minimally coupled to a scalar field $\phi$, and show that it has a single physical degree of freedom. From the original 4D perspective, the study of perturbations within the Hadad-Zakharov ansatz, using SVT decomposition, reveals that this degree of freedom corresponds to the $+$ polarised gravitational waves, while the $\times$ polarised modes are absent.

In the weakly non-linear limit, and under the eikonal approximation (equivalent to a wave-turbulence formalism), we demonstrate that imposing the set of four Einstein equations (that we solve numerically in this paper) ensures that the all  Einstein equations are satisfied for the Hadad–Zakharov spacetime. Outside these assumptions, however, this is no longer true: the Bianchi identities do not guarantee that solving these four equations implies the satisfaction of all Einstein equations for the HZ metric.

Moreover, we demonstrated that direct numerical simulations do reproduce key statistical features predicted by wave turbulence theory, including the expected Kolmogorov-Zakharov spectrum for the inverse cascade.

Nevertheless, the pseudo-spectral method remains subject to inherent limitations when applied to this type of problem. The major issue is the manifest non-conservation of the Hamiltonian constraint given by the equation~\eqref{eq:G_zz_3}, as well as the violation of the other two Einstein equations (\ref{eq:G_yy_3})-(\ref{eq:G_xx_3}), although the latter is less prominent. The origin of the problem may arise from the integration of the constraint equations~\eqref{eq:constraint_alpha_3}, \eqref{eq:constraint_beta_3} and \eqref{eq:constraint_gamma_3}, which enforce the cancellation of dynamics in the $k_x = 0$ and $k_y = 0$ modes, despite the fact that such modes are generally present. It is worth noting that the $k=0$ mode is not described in the weak turbulence regime, as the associated wave evolution time is infinite. As is often the case, this regime is therefore only valid in a limited domain of Fourier space \cite{nazarenko_2011, galtier_2022}.
Furthermore, finite box size and dealiasing may affect resolution, as evidenced by the slow propagation of the inverse front.

We would like to stress, however, that although we observe the violation of the Einstein equations that we do not solve explicitly, the Kretschmann scalar remains much larger in magnitude than the squared Ricci scalar. This suggests that the obtained results can still be regarded as a reasonably good approximation to general relativity.
We believe that, in future work, it would be appropriate to consider alternative integration methods for this class of problems to eliminate the above-mentioned inconsistencies.

Another limitation of the numerical simulation is the four-wave interaction dynamics, which require resolving time scales on the order of $\varepsilon^{-4} t_\mathrm{GW}$, thereby necessitating prohibitively large numbers of iterations. Substantial progress has been enabled by GPUs, which have significantly reduced computational time and improved statistical convergence. Allowing us to get an insight into the statistical properties of this turbulence. In particular, we found that the statistical property of the canonical variable tends to a Gaussian distribution for small-amplitude events. Still, relatively large-amplitude events occur more frequently, demonstrating intermittent behaviour. 

The analysis of intermittency, as proposed in this study, would benefit from being conducted in a statistically steady regime in which external forcing compensates for energy losses at both large and small scales. However, the slow propagation of the cascades and the limited dissipation observed in our simulations point towards a quasi-stationary regime, thereby supporting the validity of our approach.

Moreover, although current theoretical tools remain insufficient for fully characterising the statistical behaviour of the metric components, our results open new avenues for further numerical and analytical investigations. In particular, the small-scale behaviour of the gauge invariant part of the metric is rather typical in (wave) turbulence and may be an indication for the possibility of finding an exact law upon the metric such as the fourth-fifth law derived in hydrodynamics \cite{kolmogorov_41, frisch_95}. 

Several open questions arise from this work. Most notably, would the same results be recovered in a statistically forced stationary regime? 
The simulation suggests that the direct energy cascade could be driven by non-local interactions, since the Kolmogorov-Zakharov spectrum is not observed. It would therefore be interesting to study the question of locality.
Another important direction is the numerical exploration of strong turbulence. Given that our approach relies on a third-order truncation of the Einstein equations—retaining only quartic wave interactions—it is natural to ask how the cascade dynamics and associated statistics would be affected by higher-order contributions. Finally, it remains to be determined whether these findings persist within the full framework of General Relativity, beyond the specific Hadad–Zakharov reduction employed here.

\begin{acknowledgments}

BG and SG are supported by the Simons Foundation (Grant No. 651461, PPC). The work of EB and KN was supported by ANR grant StronG (No. ANR-22-CE31-0015-01).

\end{acknowledgments}

\appendix
\begin{widetext}
\section{Proof of the equivalence in the wave turbulence formalism}
\label{app:equivalence_wave_turbulence}

The use of the diagonal equations~(\ref{eq:G_zz_3})-(\ref{eq:G_xx_3}) in Fourier space leads to the following  expressions,

\begin{subequations}
\begin{align}
    \label{eq:alpha_diag}
    \tilde{\alpha}(\mathbf{k}) &= 
    - \frac{1}{4} \int_{\mathbb{R}^4} \sum_{s_1, s_2} 
    \frac{1}{p^2 \sqrt{k_1 k_2}} 
    a^{s_1}_{\mathbf{k}_1} a^{s_2}_{\mathbf{k}_2}
    e^{i \Omega_{1 2} t} \delta^0_{12}(\mathbf{k}) \\\notag
    & \hspace{75pt} \times
    \bigg[ 
        \left(\mathbf{k}_1 \cdot \mathbf{k}_2 + s_1 s_2 k_1 k_2 \right) 
        + \frac{k^2 (p_1 p_2 - q_1 q_2)}{(s_1 k_1 + s_2 k_2)^2} 
        + \frac{(p^2 - q^2) (s_1 s_2 k_1 k_2)}{(s_1 k_1 + s_2 k_2)^2} 
    \bigg] 
    ~\mathrm{d}^2\mathbf{k}_1 \mathrm{d}^2\mathbf{k}_2, \\
	\label{eq:beta_diag}
	\tilde{\beta}(\mathbf{k}) &= 
    - \frac{1}{4} \int_{\mathbb{R}^4} \sum_{s_1, s_2}
    \frac{1}{q^2 \sqrt{k_1 k_2}} 
    a^{s_1}_{\mathbf{k}_1} a^{s_2}_{\mathbf{k}_2}
    e^{i \Omega_{1 2} t} \delta^0_{12}(\mathbf{k}) \\\notag
    & \hspace{75pt} \times
    \bigg[ 
        \left(\mathbf{k}_1 \cdot \mathbf{k}_2 + s_1 s_2 k_1 k_2 \right) 
        - \frac{k^2 (p_1 p_2 - q_1 q_2)}{(s_1 k_1 + s_2 k_2)^2} 
        - \frac{(p^2 - q^2) (s_1 s_2 k_1 k_2)}{(s_1 k_1 + s_2 k_2)^2} 
    \bigg] 
    ~\mathrm{d}^2\mathbf{k}_1 \mathrm{d}^2\mathbf{k}_2, \\
    \label{eq:gamma_diag}
	\tilde{\gamma}(\mathbf{k}) &= 
    - \frac{1}{4} \int_{\mathbb{R}^4} \sum_{s_1, s_2}
    \frac{(s_1 k_1 + s_2 k_2)^2}{p^2 q^2 \sqrt{k_1 k_2}} 
    a^{s_1}_{\mathbf{k}_1} a^{s_2}_{\mathbf{k}_2}
    e^{i \Omega_{1 2} t} \delta^0_{12}(\mathbf{k}) \\\notag
    & \hspace{75pt} \times
    \bigg[ 
        \left(\mathbf{k}_1 \cdot \mathbf{k}_2 + s_1 s_2 k_1 k_2 \right) 
        - \frac{(p^2 - q^2) (p_1 p_2 - q_1 q_2)}{(s_1 k_1 + s_2 k_2)^2} 
        - \frac{k^2 (s_1 s_2 k_1 k_2)}{(s_1 k_1 + s_2 k_2)^2} 
    \bigg] 
    ~\mathrm{d}^2\mathbf{k}_1 \mathrm{d}^2\mathbf{k}_2,
\end{align}
\end{subequations}
where we used the shorthand notations $\delta^0_{12}(\boldsymbol{k})$, $\Omega^{0}_{12} = s k - s_1 k_1 - s_2 k_2$ and $\boldsymbol{k}_i = (p_i, q_i)$.

In order to assert the equivalence between the equations~(\ref{eq:alpha}--\ref{eq:gamma}) and (\ref{eq:alpha_diag}--\ref{eq:gamma_diag}), we have to compare the integrands assuming that $\boldsymbol{k} = \boldsymbol{k}_1 + \boldsymbol{k}_2$.

Let us start with the case of $\tilde{\alpha}(\boldsymbol{k})$, we have to compare:
\begin{equation}
	A = \frac{s_1 k_1 p_2 + s_2 k_2 p_1}{p (s_1 k_1 + s_2 k_2)}
    \quad
    \text{with}
    \quad
    B = 
    \frac{1}{2 p^2} 
    \bigg[ 
        \left(\mathbf{k}_1 \cdot \mathbf{k}_2 + s_1 s_2 k_1 k_2 \right) 
        + \frac{k^2 (p_1 p_2 - q_1 q_2)}{(s_1 k_1 + s_2 k_2)^2} 
        + \frac{(p^2 - q^2) (s_1 s_2 k_1 k_2)}{(s_1 k_1 + s_2 k_2)^2} 
    \bigg].
\end{equation}   
We rewrite $B$ as follows:
\begin{equation*}
\begin{aligned}
    B 
    & =
    \frac{1}{2 p^2(s_1 k_1 + s_2 k_2)^2}
    \bigg[ 
        \left(\mathbf{k}_1 \cdot \mathbf{k}_2 + s_1 s_2 k_1 k_2 \right) (s_1 k_1 + s_2 k_2)^2
        + k^2 (p_1 p_2 - q_1 q_2)
        + (p^2 - q^2) (s_1 s_2 k_1 k_2)
    \bigg] \\
    & =
    \frac{1}{2 p^2(s_1 k_1 + s_2 k_2)^2}
    \bigg[ 
        s_1 s_2 k_1 k_2 \left( p^2 - q^2 + k_1^2 + k_2^2 + 2 \mathbf{k}_1 \cdot \mathbf{k}_2 \right) 
        + \left(k_1^2 + k_2^2 \right) \mathbf{k}_1 \cdot \mathbf{k}_2
        + 2 k_1^2 k_2^2 
        + k^2 (p_1 p_2 - q_1 q_2)
    \bigg] \\
    & =
    \frac{1}{2 p^2(s_1 k_1 + s_2 k_2)^2}
    \bigg[ 
        2 s_1 s_2 k_1 k_2 \left( p_1^2 + p_2^2 + 4 p_1 p_2 \right)
        + 2 \left( p_1 + p_2 \right) \left( p_1^2 p_2 + p_2 q_1^2 + p_1 p_2^2 + p_1 q_2^2 \right)
    \bigg] \\
    & =
    \frac{1}{p (s_1 k_1 + s_2 k_2)^2}
    \bigg[ 
        s_1 s_2 k_1 k_2 \left( p_1 + p_2 \right)
        + \left( p_1 k_2^2 + p_2 k_1^2 \right)
    \bigg] \\
    & = 
    \frac{\left( s_1 k_1 + s_2 k_2 \right) \left( s_1 k_1 p_2 + s_2 k_2 p_1 \right)}{p (s_1 k_1 + s_2 k_2)^2} \\
    &= A.
\end{aligned}
\end{equation*}

Similarly, for $\tilde{\beta}(\mathbf{k})$, we need to compare:
\begin{equation}
	A = \frac{s_1 k_1 q_2 + s_2 k_2 q_1}{q (s_1 k_1 + s_2 k_2)}
    \quad
    \text{with}
    \quad
    B = 
    \frac{1}{2 q^2} 
    \bigg[ 
        \left(\mathbf{k}_1 \cdot \mathbf{k}_2 + s_1 s_2 k_1 k_2 \right) 
        - \frac{k^2 (p_1 p_2 - q_1 q_2)}{(s_1 k_1 + s_2 k_2)^2} 
        - \frac{(p^2 - q^2) (s_1 s_2 k_1 k_2)}{(s_1 k_1 + s_2 k_2)^2} 
    \bigg].
\end{equation}   
We rewrite $B$ as follows:
\begin{equation*}
\begin{aligned}
    B 
    & =
    \frac{1}{2 q^2(s_1 k_1 + s_2 k_2)^2}
    \bigg[ 
        \left(\mathbf{k}_1 \cdot \mathbf{k}_2 + s_1 s_2 k_1 k_2 \right) (s_1 k_1 + s_2 k_2)^2
        - k^2 (p_1 p_2 - q_1 q_2)
        - (p^2 - q^2) (s_1 s_2 k_1 k_2)
    \bigg] \\
    & =
    \frac{1}{2 q^2(s_1 k_1 + s_2 k_2)^2}
    \bigg[ 
        s_1 s_2 k_1 k_2 \left( q^2 - p^2 + k_1^2 + k_2^2 + 2 \mathbf{k}_1 \cdot \mathbf{k}_2 \right) 
        + \left(k_1^2 + k_2^2 \right) \mathbf{k}_1 \cdot \mathbf{k}_2
        + 2 k_1^2 k_2^2 
        - k^2 (p_1 p_2 - q_1 q_2)
    \bigg] \\
    & =
    \frac{1}{2 q^2(s_1 k_1 + s_2 k_2)^2}
    \bigg[ 
        2 s_1 s_2 k_1 k_2 \left( q_1^2 + q_2^2 + 4 q_1 q_2 \right)
        + 2 \left( q_1 + q_2 \right) \left( p_1^2 q_2 + q_2 q_1^2 + q_1 p_2^2 + q_1 q_2^2 \right)
    \bigg] \\
    & =
    \frac{1}{q (s_1 k_1 + s_2 k_2)^2}
    \bigg[ 
        s_1 s_2 k_1 k_2 \left( q_1 + q_2 \right)
        + \left( q_1 k_2^2 + q_2 k_1^2 \right)
    \bigg] \\
    & = 
    \frac{\left( s_1 k_1 + s_2 k_2 \right) \left( s_1 k_1 q_2 + s_2 k_2 q_1 \right)}{q (s_1 k_1 + s_2 k_2)^2} \\
    &= A.
\end{aligned}
\end{equation*}

Finally, for $\tilde{\gamma}(\boldsymbol{k})$, we also have to compare:
\begin{equation}
	A = \frac{p_1 q_2 + p_2 q_1}{p q}
    \quad
    \text{with}
    \quad
    B = 
    \frac{(s_1 k_1 + s_2 k_2)^2}{2 p^2 q^2} 
    \bigg[ 
        \left(\mathbf{k}_1 \cdot \mathbf{k}_2 + s_1 s_2 k_1 k_2 \right) 
        - \frac{(p^2 - q^2) (p_1 p_2 - q_1 q_2)}{(s_1 k_1 + s_2 k_2)^2} 
        - \frac{k^2 (s_1 s_2 k_1 k_2)}{(s_1 k_1 + s_2 k_2)^2} 
    \bigg].
\end{equation}   
Once again, we rewrite $B$ as follows:
\begin{equation*}
\begin{aligned}
    B 
    & =
    \frac{1}{2 p^2 q^2}
    \bigg[ 
        \left(\mathbf{k}_1 \cdot \mathbf{k}_2 + s_1 s_2 k_1 k_2 \right) (s_1 k_1 + s_2 k_2)^2
        - (p^2 - q^2) (p_1 p_2 - q_1 q_2)
        - k^2 (s_1 s_2 k_1 k_2)
    \bigg] \\
    & =
    \frac{1}{2 p^2 q^2}
    \bigg[
        \left(\mathbf{k}_1 \cdot \mathbf{k}_2 \right) (k_1^2 + k_2^2) 
        + 2 k_1^2 k_2^2
        - ((p_1 + p_2)^2 - (q_1 + q_2)) (p_1 p_2 - q_1 q_2)
    \bigg] \\
    & =
    \frac{1}{2 p^2 q^2}
    \bigg[ 
        2 s_1 s_2 k_1 k_2 \left( q_1^2 + q_2^2 + 4 q_1 q_2 \right)
        + 2 \left( q_1 + q_2 \right) \left( p_1^2 q_2 + q_2 q_1^2 + q_1 p_2^2 + q_1 q_2^2 \right)
    \bigg] \\
    & =
    \frac{1}{2 p^2 q^2}
    \bigg[ 
        \left( p_1 p_2 + q_1 q_2 \right) \left( p_1^2 + q_1^2 + p_2^2 + q_2^2 \right)
        + 2 \left( p_1^2 + q_1^2 \right) \left( p_2^2 + q_2^2 \right)
        - ((p_1 + p_2)^2 - (q_1 + q_2)) (p_1 p_2 - q_1 q_2)
    \bigg] \\
    & =
    \frac{1}{2 p^2 q^2}
    \bigg[ 
        2 p_1 p_2 (q_1^2 + q_2^2)
        + 2 q_1 q_2 (p_1^2 + p_2^2)
        + 2 p_1^2 p_2^2
        + 2 q_1^2 q_2^2
        + 4 p_1 p_2 q_1 q_2
    \bigg] \\
    & = 
    \frac{\left( p_1 + p_2 \right) \left( q_1 + q_2 \right)}{\left( p_1 + p_2 \right)^2 \left( q_1 + q_2 \right)^2} \left( p_1 q_2 + p_2 q_1 \right) \\
    &= A.
\end{aligned}
\end{equation*}
Finally, since all integrands are the same, we conclude the equivalence (in the wave turbulence paradigm) of the diagonal and the off-diagonal equations.

\end{widetext}

\bibliography{bibliography}

\end{document}